# How Grain Boundaries and Interfacial Electrostatic Interactions Modulate Water Desalination Via Nanoporous Hexagonal Boron Nitride

*Bharat Bhushan Sharma and Ananth Govind Rajan*[*]

Department of Chemical Engineering, Indian Institute of Science, Bengaluru, Karnataka 560012, India

**\*Corresponding Author:** Ananth Govind Rajan (Email: ananthgr@iisc.ac.in)

**ABSTRACT**

To fulfil the increasing demand for drinking water, researchers are currently exploring nanoporous two-dimensional materials, such as hexagonal boron nitride (hBN), as potential desalination membranes. A prominent, yet unsolved challenge is to understand how such membranes will perform in the presence of defects or surface charge in the membrane material. In this work, we study the effect of grain boundaries (GBs) and interfacial electrostatic interactions on the desalination performance of bicrystalline nanoporous hBN, using classical molecular dynamics simulations supported by quantum-mechanical density functional theory (DFT) calculations. We investigate three different nanoporous bicrystalline hBN configurations, with symmetric tilt GBs having misorientation angles of 13.2°, 21.8°, and 32.2°. Using lattice dynamics calculations, we find that grain boundaries alter the areas and shapes of nanopores in bicrystalline hBN, as compared to the nanopores in monocrystalline hBN. We observe that, although bicrystalline nanoporous hBN with a misorientation angle of 13.2° shows improved water flow rate by ~30%, it demonstrates reduced $Na^+$ ion rejection by ~6%, as compared to monocrystalline hBN. We also uncover the role of the nanopore shape in water desalination, finding that more elongated pores with smaller sizes (in 21.8°- and 32.2°-misoriented bicrystalline hBN) can match the water permeation through less elongated pores of slightly larger sizes, with a concomitant ~3-4% drop in $Na^+$ rejection. Simulations also predict that the water flow rate is significantly affected by interfacial electrostatic interactions. Indeed, the water flow rate is the highest when altered partial charges on B and N atoms were determined using DFT calculations, as compared to when no partial charges or bulk partial charges (i.e., charged hBN) were considered. Overall, our work on water/ion transport through nanopores in bicrystalline hBN indicates that the presence of GBs and surface charge can lead, respectively, to a drop in the ion rejection and water permeation performance of hBN membranes.

**INTRODUCTION**

Increased population, global warming, and industrialization have all put pressure on the supply of fresh and potable water.[1–3] Today, about 15% of the world's population faces challenges in obtaining drinking water.[4,5] Although an abundant amount of water is present on the earth's surface, only 3% of it is available as potable water.[1,6] In contrast, the remaining 97% of water is present as salty water in the oceans and seas.[1,6] Consequently, the desalination of seawater is a promising solution to satisfy the demand for fresh and potable water, in the near and long-term future. Although numerous methods are available for the desalination of seawater, membrane-based reverse osmosis (RO) is extensively used at both commercial and domestic levels due to its higher energy efficiency.[7] The performance of a RO system depends significantly on the filtration membrane employed. Thus, the role played by the membrane material in a RO system is very crucial.[7] RO systems traditionally use ceramic and polymeric membranes, which face limitations such as lower strength, water permeability, and service life, and higher power requirements and maintenance cost.[8] In view of these limitations of conventional membranes, researchers are currently exploring nanoporous two-dimensional (2D) materials such as hexagonal boron nitride (hBN) and graphene as desalination membranes.[6,9–12] As opposed to graphene, hBN and molybdenum disulfide ($MoS_2$) membranes provide opportunities for creating anion- and cation-selective membranes, as boron and nitrogen atoms in hBN, and molybdenum and sulfur atoms in $MoS_2$, are positively and negatively charged, respectively, and can be used to terminate the nanopores.[10,13] However, the smaller lateral dimensions of 2D materials obtained using various synthesis techniques poses a challenge in terms of scaling up the process.[6,11,14,15]

Indeed, for commercial water desalination, large-area 2D materials are required to obtain adequate production rates of potable water. Typically, extended 2D layers are produced by the chemical vapor deposition (CVD) method, which inevitably introduces grain boundaries (GBs) in the material due to temperature gradients in the reactor and multiple nucleation sites on the growth surface.[16] Although the monocrystalline forms of nanoporous hBN and graphene nanosheets have extraordinary desalination performance,[6,17,18] the effect of GBs on the water permeability and ionic rejection of such membranes is unknown. Nevertheless, it is well-known that inadvertently induced GBs in these 2D nanosheets alter their mechanical performance and properties.[19] Furthermore, quantification of the effect of defects on water/ion transport through single-digit nanopores is a prominent knowledge gap in the field.[20] Over the years, classical molecular dynamics (MD) simulations and quantum-mechanical density functional theory (DFT) calculations have been used to understand and predict the atomic-level phenomena underlying water permeation through and ion rejection by nanoporous 2D materials.[6,10,17,21–23] However, so far, simulation studies have considered idealized models of nanoporous hBN and graphene that do not contain GBs in them.

In 2015, Surwade et al.[9] experimentally analyzed the capabilities of nanoporous graphene as a desalination membrane. In their work, the oxygen plasma etching method was used to create nanopores in single-layer graphene. They concluded that the resultant nanoporous graphene membranes exhibit excellent water permeability and 100% salt rejection. Apart from this experimental study, several modeling studies have been carried out to analyze the desalination performance of nanoporous 2D materials.[6,21–29] In 2012, Cohen-Tanugi et al.[6] carried out classical MD simulations and concluded that nanoporous graphene shows extraordinary water permeability, which is many times larger than conventional membranes. They also reported that the water permeability depends on the nanopore's size, edge functionalization, and applied pressure. Subsequently, Konatham et al.[24] performed MD simulations to analyze the water desalination performance of graphene membranes and reported that nanopores of diameter ~7.5 Å exhibit effective ion segregation, whereas ions easily passed through nanopores with diameter between ~ 10.5 Å and 14.5 Å. These studies uncovered the critical role of the nanopore size in modulating water and ion transport through 2D material membranes. In addition to graphene, other 2D materials have also been examined. Heiranian et al.[23] performed MD simulations to investigate the desalination performance of single-layer nanoporous $MoS_2$ and predicted that nanopores with molybdenum atoms on their edges showed higher water fluxes (~70% greater than graphene nanopores). Recently, a review on $MoS_2$-based membranes in water treatment and purification was published wherein the authors summarized several aspects of $MoS_2$ membranes including surface modification.[30] In another study, Cao et al.[22] analyzed the desalination performance of 2D metal organic framework (MOF) membranes using MD simulations and concluded that 2D MOF membranes lead to 3-6 times higher water permeation than traditional membranes. It was also reported that few-layered MOF membranes exhibit higher water flux than single-layer graphene or $MoS_2$ membranes without any drilling of nanopores, due to the inherent pore structure of the material.

Besides the above-mentioned 2D nanosheets, hBN nanosheets have also been investigated as separation membranes by various researchers.[4,10,35–37,12,15,17,18,31–34] Chen et al.[12] experimentally examined the water transport performance of amino-functionalized nanoporous hBN using a positive pressure setup. They found that the hBN membranes are highly stable in various organic solvents and water and showed high water flux and excellent selectivity in organic and aqueous solvents. In other experimental work, Lei et al.[38] used a thermal treatment method to synthesize nanoporous hBN and demonstrated its potential to purify contaminated water by absorbing various solvents, dyes, and oils. With respect to modeling studies, Garnier et al.[17] carried out MD simulations to predict the surface tension profile of water molecules close to mono and multi-layer hBN and graphene nanosheets and to investigate water permeability through nanoporous hBN and graphene. The authors established a correlation between the surface tension profile and the water permeation rate. They further reported that the water surface tension was lower on monolayer hBN as compared to monolayer graphene, which resulted in faster water permeation through monolayer nanoporous hBN due to increased wetting. Gao et al.[18] employed classical MD simulations to determine the desalination performance of hBN nanosheets and investigated how the shape and size of nanopores in hBN affected salt rejection and water permeation. They also showed that triangular nanopores with nitrogen atoms on the edges of the nanopore exhibited high water permeation rates in the range of 0.132 to 2.752 kg m$^{-2}$ s$^{-1}$ MPa$^{-1}$ which is several times higher than the water permeation rates through commercial RO membranes (~ 0.011 kg m$^{-2}$ s$^{-1}$ MPa$^{-1}$). Davoy et al.[15] used MD simulations to predict that the water permeability of nanoporous hBN far exceeded that of polymer-based conventional RO membranes. In fact, it was even higher than that through nanoporous graphene. Other studies focused on unravelling the role of doping,[4] nanopore edge functionalization,[10,35,37] and edge type[33] in modulating water permeation, have also been carried out. Nevertheless, in the above mentioned studies, the electrical potential produced by nanoporous hBN was not validated with quantum-mechanical DFT calculations, although the charges on B and N atoms were derived using some DFT-based charge partitioning scheme. In fact, in some studies partial charge values from bulk hBN were used, which may lead to underprediction of the water flux. This aspect is crucial because, as we show, electrostatic interactions play a key role in modulating the performance of nanoporous hBN in desalination applications.

From previous studies, it is evident that monocrystalline nanoporous hBN without any geometrical defects exhibits excellent water permeability[17], adsorption ability[39], mechanical strength[19], and structural stability.[12,19,40] Nevertheless, as mentioned before, for water desalination applications, large-area 2D materials are required, which would lead to the presence of grain boundaries (GBs) in the membrane. So far, studies have investigated the water desalination performance of nanoporous monocrystalline hBN and no study has considered the effect of GBs on the desalination performance of nanoporous hBN. In this work, we investigate the role of GBs in modulating the desalination performance of bicrystalline hBN using classical

MD simulations. Because GBs and nanopores alter the charge distribution within the hBN nanosheet, thereby affecting electrostatic interactions between hBN and water, we use quantum-mechanical DFT to obtain accurate partial charges on the B and N atoms. We also show that the resultant electrostatic interactions significantly affect the desalination performance of nanoporous bicrystalline hBN. Finally, we conclude that the presence of GBs can adversely affect the desalination performance of nanoporous hBN, due to a significant drop in $Na^+$ ion rejection through the membrane.

**METHODS**

*Computational modeling and simulation details.* We used classical MD simulations to study the desalination performance of nanoporous bicrystalline hBN using the open-source Large-scale Atomic/Molecular Massively Parallel Simulator (LAMMPS) package.[41] The simulation system for water desalination is illustrated in Figure 1, in which nanoporous hBN was placed parallel to the *x-y* plane. We used a simulation box measuring 160 Å in the *z*-direction and periodic in the *x-y* directions with a cross-sectional area of ~31 × 31 Å. The nanoporous hBN sheet was fixed at $z = 0$ Å, and rigid graphene pistons were placed at $z = -75$ and $z = 18.5$ Å. The feed side region of the simulation box was filled with saline water having 22 $Na^+$ ions, 22 $Cl^-$ ions, and 2000 water molecules, leading to a salt concentration of 35.7 gm/L, close to the value observed in seawater. The permeate (i.e., pure water) side was filled with 500 water molecules. We verified that the use of a larger number of water molecules on the permeate side did not affect our results. Indeed, the water flow rates through the nanoporous monocrystalline hBN membrane at 50 MPa feed pressure were found to be the same when 500 and 2000 water molecules were present on permeate side, as shown in Figure S1 in Supporting Information Section S1.

In all the simulations, the atoms in the nanoporous hBN membrane were allowed to vibrate, except for the ones closest to the periodic boundary. The latter atoms were kept fixed to maintain the hBN sheet at $z = 0$ Å by zeroing their momentum. Periodic boundary conditions were implemented along all the three directions. All simulations were carried out at room temperature (298.15 K) and atmospheric pressure (1 atm) in the lateral, i.e., $x$-$y$ direction, using the $NP_{xy}T$ ensemble. These conditions were maintained using the Nose-Hoover thermostat and barostat.[42,43]

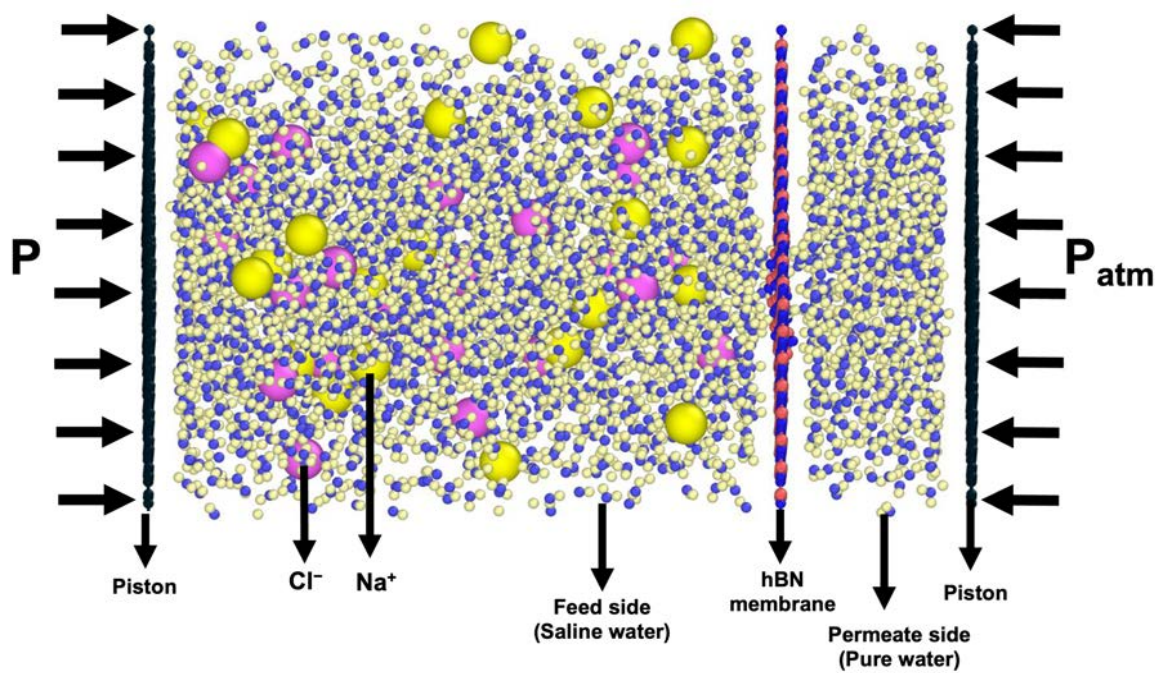

**Figure 1.** Schematic of the water desalination simulation setup confined by graphene pistons. The saline water reservoir (feed side, left) and pure water reservoir (permeate side, right) are separated by a nanoporous hBN membrane. The pressure on the feed site is $P$ and that on the permeate side is $P_{atm} = 1.013$ bar. The water molecules, sodium ions, and chloride ions are represented using spheres with hydrogen in cream, oxygen in blue, Na$^+$ in yellow, and Cl$^-$ in magenta.

After carrying out energy minimization, the system was equilibrated in the $NP_{xy}T$ ensemble for a total time period of 50 ps. A short equilibration phase was chosen so that the ions and/or water molecules do not move to the permeate side. The total energy and temperature were recorded during equilibration and are plotted in Figure S2 in Section S2. It can be deduced from Figure S2a and S2b that the total energy and temperature are converged after 50 ps, indicating that the short equilibration time chosen is sufficient. In all the simulations, a time step of 1 fs was used. LJ interactions were cut-off at a distance of 1.2 nm for water-water and hBN-water interactions. Intra-hBN short-range LJ/Coulombic interactions were not considered. Long-range electrostatic interactions were included using the particle-particle–particle-mesh (PPPM) method.[44] Post equilibration, a uniform pressure ($P$) was applied on the feed side piston along the +$z$ direction, i.e., the flow direction, whereas atmospheric pressure was implemented on the permeate side piston in the -$z$ direction, as shown in Figure 1. As explained in the Results and Discussion section, three different feed-side pressures, i.e., 50 MPa, 100 MPa, and 200 MPa were used, leading to effective pressure values of 46.87 MPa, 96.87 MPa, and 196.87 MPa, respectively, which are similar in magnitude to the values considered in previous theoretical studies.[17,18] The applied pressure was implemented in the form of a uniform force, calculated as $F = PA/n$, where $P$ denotes the applied pressure, $A$ the piston cross-sectional area, and $n$ the number of atoms in the piston.

***Details of the force fields used.*** In MD simulations, selection of the force-field parameters is very critical, as the accuracy of the simulation critically depends on them. In this work, a hybrid model (combining the Tersoff potential for hBN,[45] the TIP4P/2005 model for water,[46] and 12-6 LJ plus Coulombic interactions between hBN and water) was used to capture the atomic interactions among different atoms. The Tersoff model with parameters suggested by Albe et al.[45] was used to model the nanoporous monocrystalline and bicrystalline hBN membrane, as done previously by Loh in the context of desalination[4] and by other researchers.[47–49] The TIP4P/2005 water model[46] was used to represent water molecules, whereas Lennard-Jones (LJ)[50] and Coulombic interatomic potentials were used for all the intermolecular interactions. The parameters for LJ and Coulombic interatomic potentials were taken from Konatham et al.[24] for carbon atoms, Govind Rajan et al.[51] for B and N atoms (only LJ parameters), and Zeron et al.[52] for the sodium and chloride ions; these are tabulated in Section S3. In previous work, it was shown that the LJ parameters proposed by Govind Rajan et al. for B and N atoms[51] are able to reproduce qualitatively the water contact angle on the hBN basal plane.[53] To calculate the interactions between unlike atoms in the system, geometric mean combining rules, $\sigma_{ij} = (\sigma_i \sigma_j)^{\frac{1}{2}}$ and $\varepsilon_{ij} = (\varepsilon_i \varepsilon_j)^{\frac{1}{2}}$, were used where, $\sigma$ represents the distance at which the interatomic LJ potential is zero and $\varepsilon$ represents the LJ well-depth parameter. In Section S4, we present a more detailed discussion on the choice of force fields for hBN, water molecules, and salt ions (see, e.g., refs.[14] and [54]).

***Partial charge calculations.*** To calculate the partial charges, the geometry of monocrystalline and bicrystalline nanoporous hBN membranes was optimized using the Broyden–Fletcher–Goldfarb–Shanno (BFGS) algorithm, leading to a maximum force of 0.0003 hartree/bohr on the atoms. The DFT calculations were performed using the cp2k package[55] and the Perdew-Burke-Ernzerhof (PBE) exchange-correlation functional[56] with double zeta (short range) valence basis sets for the B and N atoms.[57] The nuclei and core electrons were modeled using the Goedecker-Teter-Hutter pseudopotentials.[58] Subsequently, partial charges on B and N atoms were calculated using the density derived atomic point (DDAP) charge algorithm developed by Blöchl.[59] After calculating the partial charges, the electrostatic potential plots in Figures 5 and S6 were prepared by calculating the electrostatic interaction between a unit test charge and the nanoporous hBN membrane (see Figure S3a in Section S5) using the group/group command in LAMMPS. The same group/group command was also used to plot the contours of the interfacial interactions between a single, tangential water molecule and the nanoporous hBN membrane in Figures 12 and 13. The simulation setup used to calculate these interactions is shown in Figure S3b in Section S5.

**RESULTS AND DISCUSSION**

***GB misorientation angles and nanopore shapes in hBN.*** In this work, we modeled water and ion permeation through four different hBN membranes: (i-iii) bicrystalline nanoporous hBN with symmetric tilt GBs having misorientation angles ($\theta$) of (i) 13.2°, (ii) 21.8°, and (iii) 32.2° and (iv) monocrystalline nanoporous hBN (i.e., without any GB), as shown in Figure 2. We considered the monocrystalline hBN membrane to compare our findings with previously reported results and to quantify the effect of GBs on the desalination performance of nanoporous hBN. The nomenclature of the GB structures considered and details regarding their repeat length are described in the Section S6. The bicrystalline hBN configurations were created using the sewing method,[60] which generates pairs of pentagon-heptagon (5|7) dislocations at the junctions of the two misoriented crystals of hBN, as shown in Figure 2. Similar types of GBs structures in hBN and graphene have been seen experimentally[61–65] and predicted theoretically[49,66–69] in various studies. We chose these three misorientation angles because they cover varying linear dislocation densities (i.e., the number of (5|7) dislocation pairs on the GB per unit length) from low to high (~0.91 nm$^{-1}$, 1.5 nm$^{-1}$, and 2.24 nm$^{-1}$).

Nanopores were considered at the GB, because it is more likely for extended defects, such as nanopores, to form there. We considered roughly triangular nanopores with nitrogen-terminated edges, because triangular, boron-deficient defects are more-likely to form in hBN, as seen experimentally[70–72] and predicted theoretically[73,74] by previous studies. For nanoporous monocrystalline hBN, we considered a nanopore formed by removing 10 B atoms and 6 N atoms from the hBN lattice, leading to a total of 16 removed atoms (Figure 2a,b). Indeed, in previous work, Govind Rajan et al.[73] cataloged the most-probable nanopore shapes in hBN using kinetic Monte Carlo simulations, DFT calculations, and graph theory and predicted triangular shapes of nanopores with nitrogen atoms at their edges to be the most prevalent in hBN. Further, Kozawa et al.[74] predicted that triangular shapes of nanopores are the most kinetically favorable defects in hBN when a perfect square number of atoms (16 here) was removed. The choice of considering a triangular nanopore in hBN is also substantiated by the experimental observation of triangular vacancy defects in hBN.[75,76] The triangular nanopore formed by removing 25 atoms (the next perfect square after 16) was not considered, because of the trade-off between an increased water flux and reduced ion rejection. Indeed, while the water flux can be improved by considering a larger nanopore, the ion rejection ability of the membrane will severely deteriorate in such a scenario. Thus, to investigate the overall efficiency of a nanoporous membrane for desalination purposes, a trade-off between water flux and ions rejection is very crucial. For the 13.2°-misoriented bicrystalline hBN, an equal number of B and N atoms as in the monocrystalline case (10 and 6, respectively) were removed to form the nanopore (Figure 2c,d); yet the area of the nanopore is slightly larger (27.6 Å$^2$ versus 26.4 Å$^2$) due the presence of the GB. The calculation of the nanopore area is explained in Section S7.

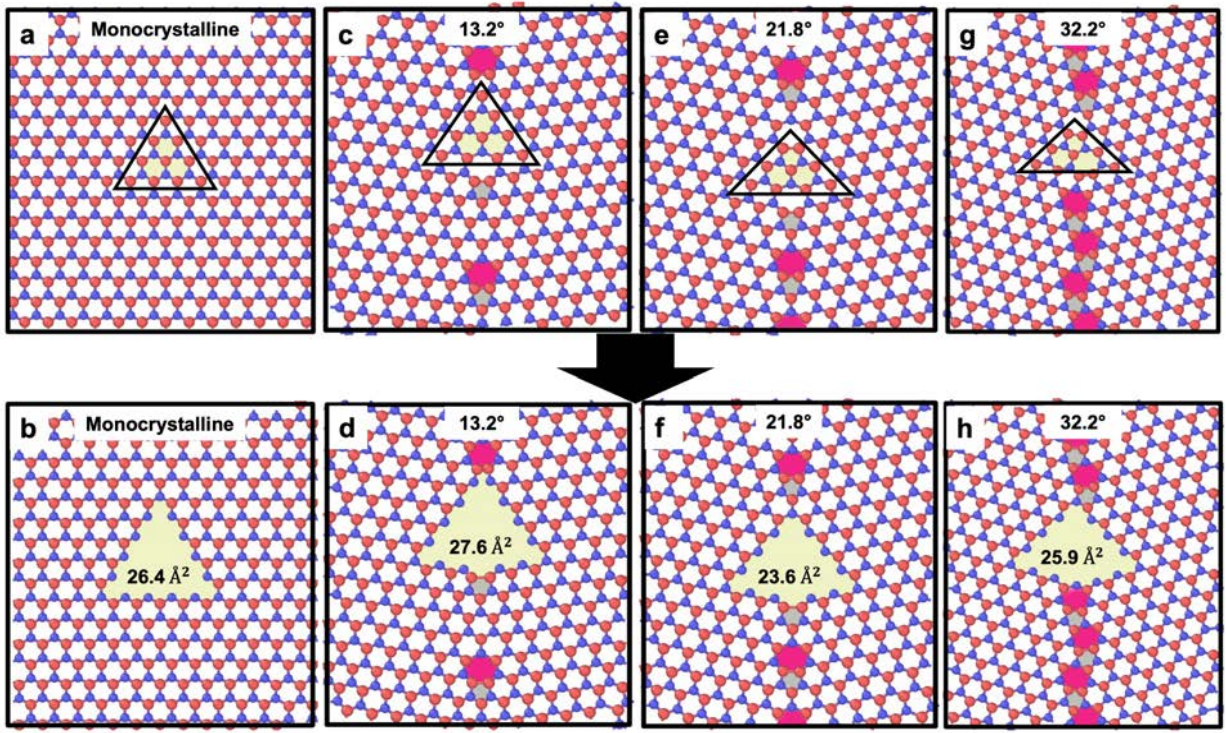

**Figure 2.** Snapshots of monocrystalline and bicrystalline hBN without (a,c,e,g) and with (b,d,f,h) a roughly triangular, N-terminated nanopore: (a-b) monocrystalline hBN, (c-d) 13.2°-misoriented bicrystalline hBN, (e-f) 21.8°-misoriented bicrystalline hBN, and (g-h) 32.2°-misoriented bicrystalline hBN. Atoms on and inside the black triangular outline are removed to create the nanopore in each case. Red and blue spheres represent boron and nitrogen atoms, respectively. The pore area is indicated in each case. The panels e and g have a smaller width than panels a and c, because a larger GB repeat length (in the vertical, i.e., $y$ direction) necessitated the use of a smaller width (in the horizontal, i.e., $x$ direction) to maintain roughly equal nanopore concentrations of ~0.09-0.10 nm$^{-2}$.

Note that, our estimate of 26.4 Å$^2$ for the area of the nanopore in monocrystalline hBN formed by removing 10 B and 6 N atoms from the hBN lattice is similar to the estimate of 29.1 Å$^2$ obtained by Davoy et al.[15] For the 21.8°-misoriented bicrystalline hBN, 9 B atoms and 5 N atoms were removed to form a roughly triangular nanopore (Figure 2e,f); the resultant nanopore area is the smallest at 23.6 Å$^2$. Finally, for the 32.2°-misoriented bicrystalline hBN, 10 B atoms and 5 N atoms were removed to form a roughly triangular nanopore (Figure 2g,h), leading to an intermediate nanopore area of 25.8 Å$^2$. It is evident that the presence of a GB affects the nanopore area and shape. Later in the article, we also discuss the water desalination performances of these nanopores after normalization by the nanopore area. In all the considered cases, the nanopore concentration (i.e., pore density) in the hBN layer was ~0.09-0.10 nm$^{-2}$ (see the caption of Figure 2). This is a large value considering that almost 10$^{17}$ nanopores would be present in 1 m$^2$ area of the membrane. Nevertheless, our estimates of the per pore permeance/permeability can simply be multiplied by the actual pore density to obtain the membrane permeance/permeability, for membranes with a lower number of pores per unit area.[77] In Table 1, we list the areas of the nanopores considered, their equivalent circular diameters, and their aspect ratios, as quantified by the ratio of the minimum to maximum Feret diameter of the nanopore (see Section S7 for more details). The Feret diameter, also known sometimes as the caliper diameter, is the distance between the jaws of a caliper used to measure the size of an object. Note that the aspect ratio, as defined above, lies between 0.0 and 1.0. The limiting value of 1.0 corresponds to a circle and that of 0.0 to a line segment; the aspect ratio of an equilateral triangle, equaling the ratio of its height to its side length, is $\frac{\sqrt{3}}{2} = 0.87$. Thus, the lower the aspect ratio of a nanopore, the more elongated it is. From Table 1, we see that the extent of elongation of the considered nanopores is in the order: monocrystalline hBN < 13.2°-misoriented hBN < 21.8°-misoriented hBN < 32.2°-misoriented hBN.

**Table 1.** Geometrical properties of the four nanopores considered in this work.

| Configuration | Area (Å$^2$) | Equivalent circular diameter (Å) | Aspect ratio (-) |
|---|---|---|---|
| **Monocrystalline** | 26.4 | 5.80 | 0.91 |
| **13.2°** | 27.6 | 5.93 | 0.84 |
| **21.8°** | 23.6 | 5.49 | 0.78 |
| **32.2°** | 25.9 | 5.74 | 0.62 |

***Partial charge calculations and validation of the resultant electrostatic potential with DFT.*** As discussed above, hBN membranes used for water desalination have intentionally induced geometrical defects such as nanopores and inadvertently induced geometrical defects such as GBs. These geometrical defects in hBN alter the partial charges on B and N atoms, ultimately affecting the electrostatic interactions between water and hBN. To quantify this effect, we calculated the partial charges on B and N atoms in the hBN membranes considered here using the DFT-based DDAP charge algorithm (see Computational Methods for more details). Specifically, partial charges were calculated using nine different sets of DDAP parameters, as described in Section S8. Here, we focus on cases 1 (the default DDAP parameters) and 4 (the "best" DDAP parameters, see below). The contours plots of the partial charges calculated using the default and best DDAP parameters are indicated in Figure 3a-d and Figure 4a-d, respectively, for the four cases (i)-(iv) discussed above. It can be inferred from Figure 3b-d and Figure 4b-d that in bicrystalline hBN membranes, B atoms adopt negative charges and N atoms adapt positive charges when DDAP parameters are altered from their default values to their best values (see for the parameter values). In contrast, no such sign reversal in the partial charges of B and N atoms takes place in a monocrystalline membrane when using the best DDAP parameters, as seen by comparing Figure 3a and Figure 4a. This sign reversal, found essential to reproduce the DFT-predicted electric potential, as discussed below, could be caused by a polarizing electric field at the GB parallel to the hBN surface and perpendicular to the GB, and should be probed more deeply in the future.

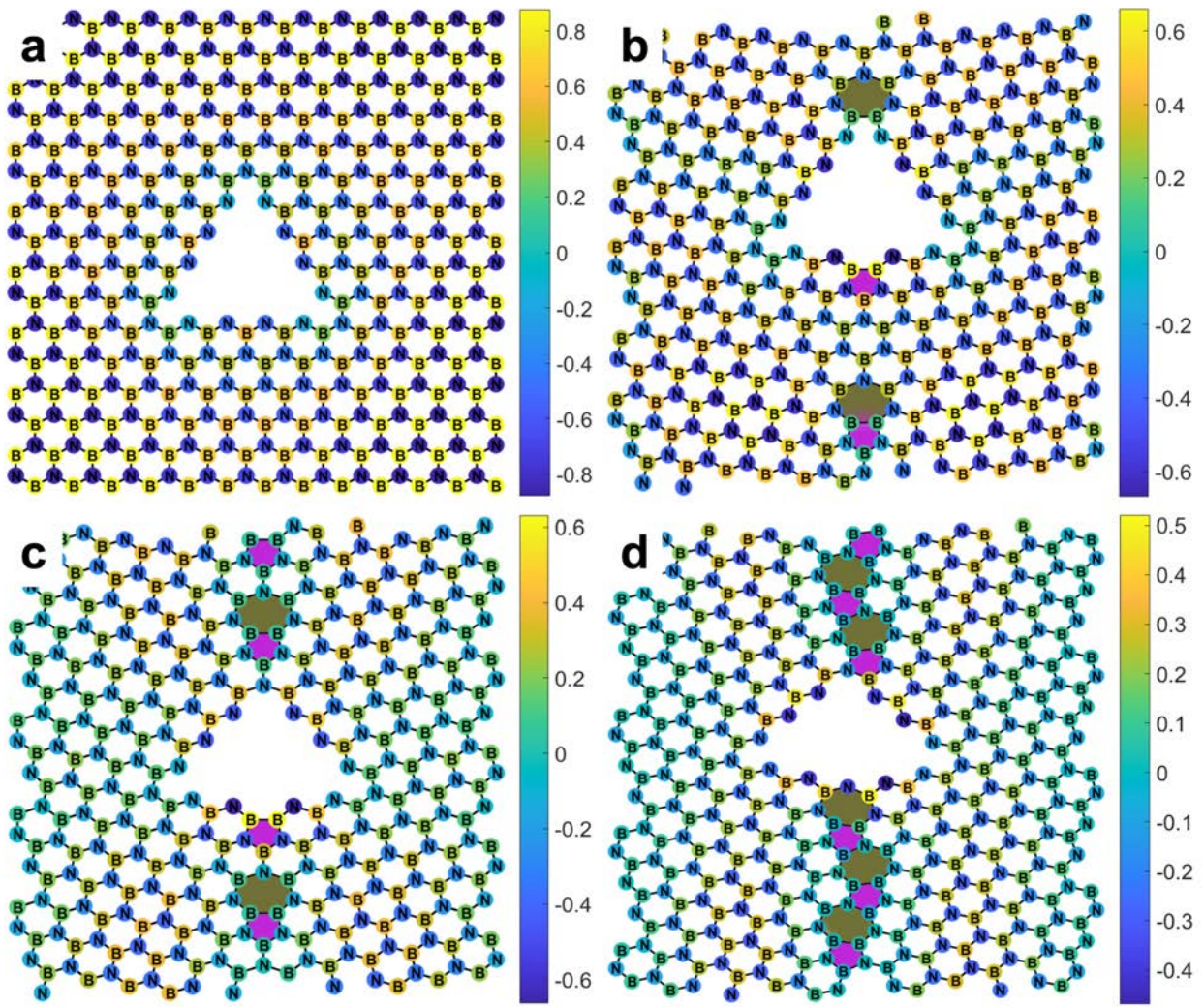

**Figure 3.** Partial charge (in elementary charge unit ($e$)) distribution calculated using the default DDAP parameters for various configurations: (a) Monocrystalline (b) 13.2°-misoriented bicrystalline, (c) 21.8°-misoriented bicrystalline, and (d) 32.2°-misoriented bicrystalline nanoporous hBN membranes, respectively. Five membered rings are colored in magenta and seven membered rings are colored in green.

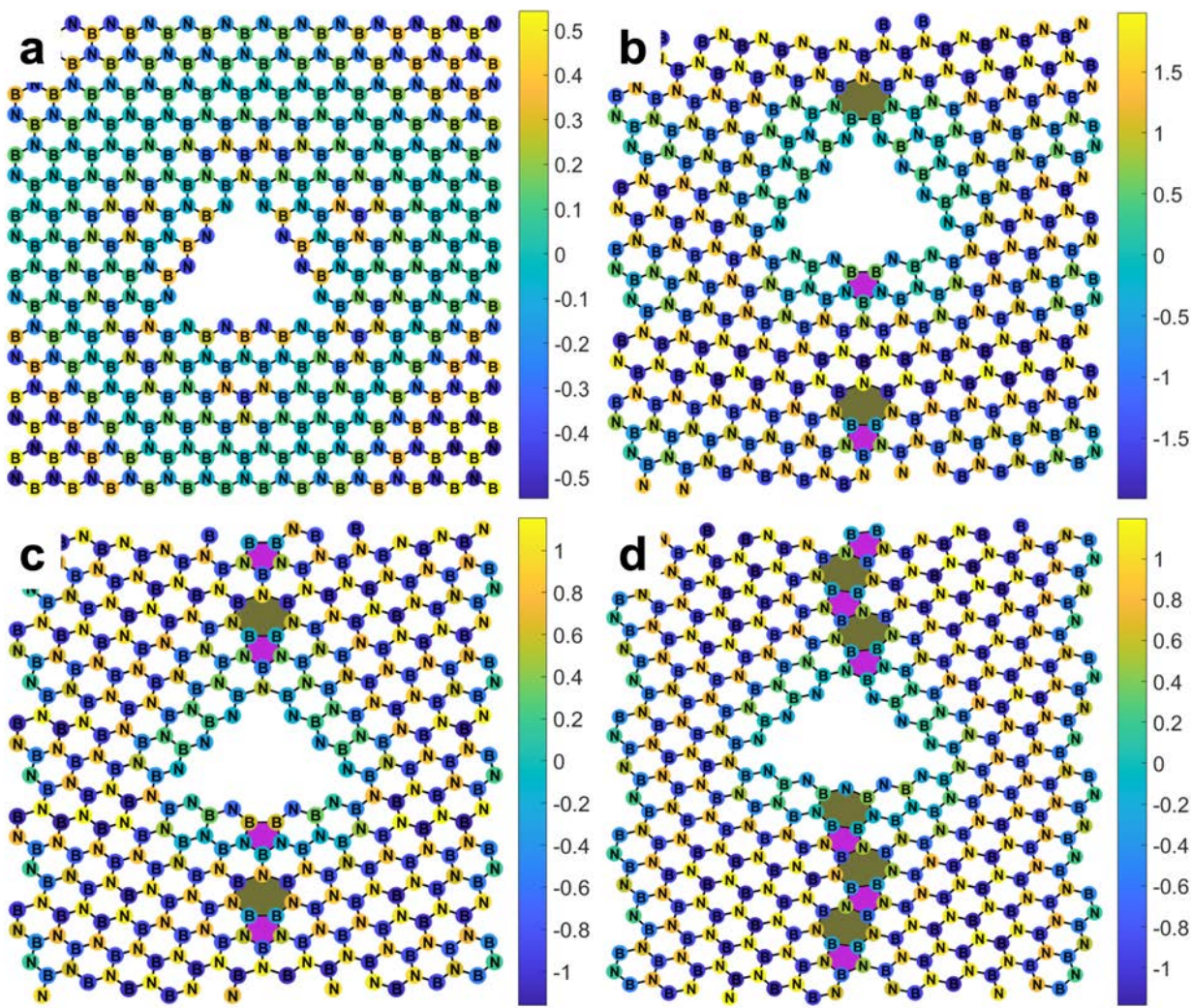

**Figure 4.** Partial charge (in elementary charge unit ($e$)) distribution calculated using the "best" DDAP parameters for various configurations: (a) Monocrystalline (b) 13.2°-misoriented bicrystalline, (c) 21.8°-misoriented bicrystalline, and (d) 32.2°-misoriented bicrystalline nanoporous hBN membranes, respectively. Five membered rings are colored in magenta and seven membered rings are colored in green.

To validate the DFT-predicted partial charges, we also calculated the electrostatic potential generated by the total DFT charge density (electrons and ions) in the hBN membrane. This was achieved by writing a cube file using the V_HARTREE_CUBE section in cp2k. The electrostatic potential values at points directly above the center of the nanopore were plotted against the perpendicular distance from the surface of the hBN sheet for both monocrystalline and bicrystalline configurations, as shown in Figure 5. To obtain the potential created by the partial charges on B and N atoms, a unit test charge was placed at the centre of the nanopore of the hBN membrane, as depicted in Section S5. This unit test charge was moved in the direction perpendicular to the hBN surface, and the electrostatic interactions between the hBN layer and the test charge were obtained using single-point calculations in the LAMMPS package.[41] The resultant electrostatic potential was compared with the DFT potential as shown in Figure 5. It can be inferred from Figure 5a-d that the electrostatic potentials calculated using the partial charges with default DDAP parameters do not agree with the DFT potential. In contrast, the electrostatic potentials calculated using the partial charges with the "best" DDAP parameters for nanoporous monocrystalline and bicrystalline (21.8°- and 32.2°-misoriented) hBN (see Figure 5a,c-d) are in excellent agreement with the DFT potential. In the case of the 13.2°-misoriented bicrystalline hBN membrane, there is a minor deviation between the DFT potential curve and the electrostatic potential curve calculated using the partial charges with the best DDAP parameters. We show in the Section S8 that it is not possible to eliminate these differences by tuning the partial charges. Nevertheless, later in the article, we show that this minor deviation between the DFT and partial-charge-based potentials does not affect the main conclusion of our work. It is also seen in Figure 5 that the electrostatic potential is the highest at the center of the pore (the hBN layer is placed at $z = 0$ Å), with a symmetric drop in potential, on either side, as one moves away from the hBN layer.

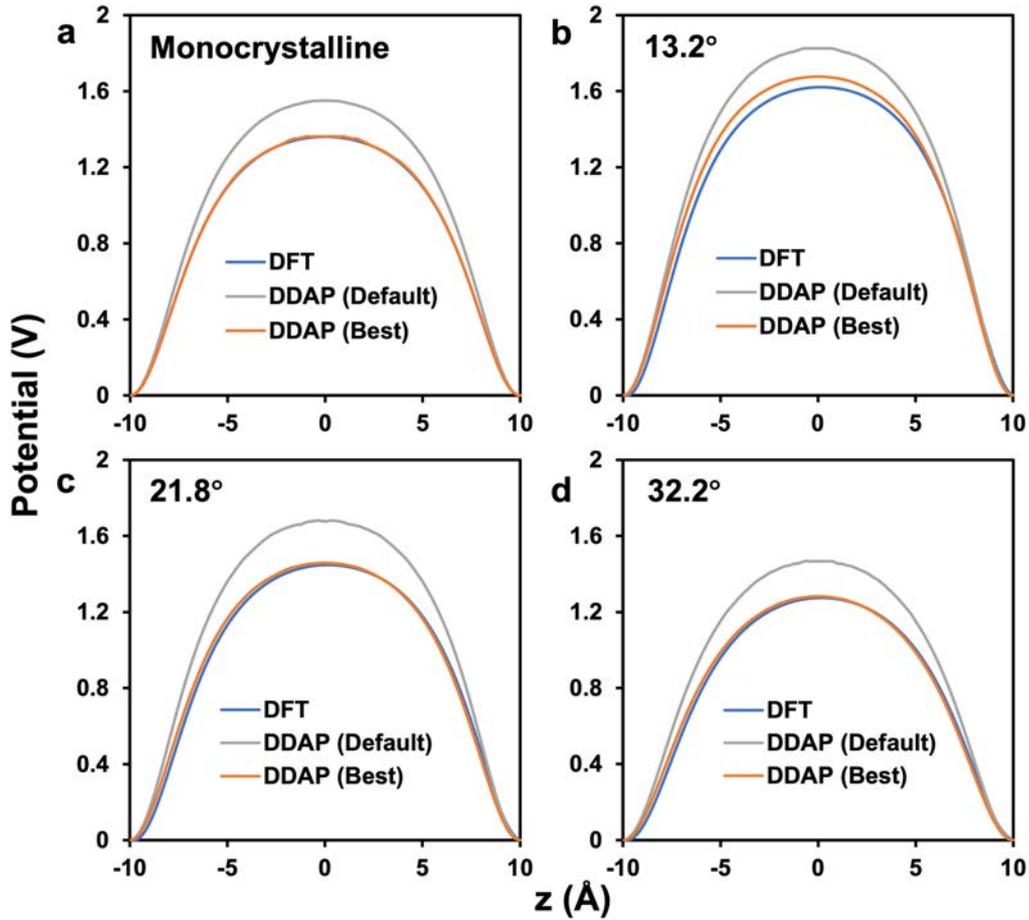

**Figure 5.** Electrostatic potential as a function of perpendicular distance from the surface of membrane for various configurations: (a) monocrystalline, (b) 13.2°-misoriented bicrystalline, (c) 21.8°-misoriented bicrystalline, and (d) 32.2°-misoriented bicrystalline nanoporous hBN membranes. Values calculated using DFT (purple), partial charges based on the default DDAP parameters (grey), and partial charges based on the best DDAP parameters (orange) are shown. The hBN membrane is placed at $z = 0$ Å in each case.

***Water flow rate through nanoporous monocrystalline and bicrystalline hBN.*** After calculating and validating the partial charges, MD simulations were carried out to investigate the desalination performance of nanoporous monocrystalline and bicrystalline hBN membranes. The simulation system for water desalination is illustrated in Figure 1 and consists of a nanoporous hBN layer, two rigid graphene pistons, water molecules, and sodium ($Na^+$) and chloride ($Cl^-$) ions. To analyze the desalination performance of these membrane configurations, the number of water molecules that permeated through the membrane ($N_w$) was counted and plotted as a function of time at different feed pressures ranging from 50 to 200

MPa in Figure 6. The permeate pressure is maintained at 0.1 MPa (i.e., 1 bar) in all cases. All simulations for water permeation were run till 200 water molecules (10% of the total water molecules on the feed side) permeated through the membrane. This is done so that the salinity of the water in the feed side is not significantly affected, thereby preventing large variations in chemical potential gradients during the water and ion permeation process. Therefore, the simulations times vary as a function of the applied pressure (as shown in Figure 6), and increase from ~2 ns to ~9 ns as the applied pressure decreases from 200 MPa to 50 MPa. Note that, in some of the previously published studies,[10,18] simulations were performed for a longer time (~30 ns), thereby significantly affecting the salinity of water on the feed side. One can observe from the plotted trend in Figure 6 that in all the membrane configurations, the number of permeating water molecules followed a linear trend with simulation time, which is a signature of a constant pressure and chemical potential driving force. Moreover, the number of permeating water molecules significantly depended upon the imposed pressure. As the imposed pressure increased, the value of $N_w$ at fixed time also increased for all the hBN configurations.

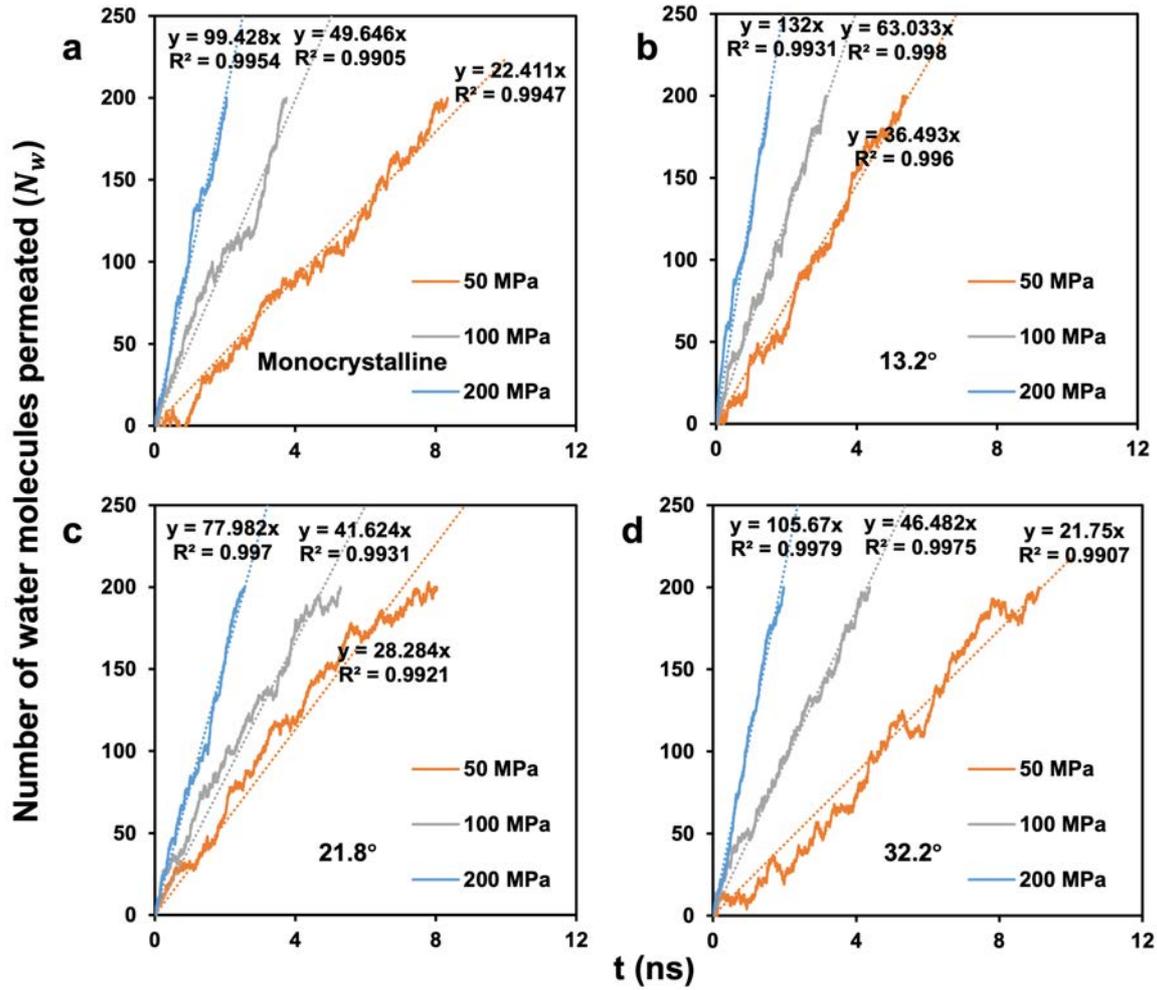

**Figure 6.** Plots of the number of water molecules that permeated through the nanoporous hBN membrane versus time for various configurations: (a) monocrystalline, (b) 13.2°-misoriented bicrystalline, (c) 21.8°-misoriented bicrystalline, and (d) 32.2°-misoriented bicrystalline hBN. The plots are shown at three different feed pressures: 50 MPa (orange), 100 MPa (gray), and 200 MPa (purple), with dotted lines showing linear fits. The slope and the goodness of fit ($R^2$ value) are indicated in each case.

After plotting $N_w$ versus time, we calculated the water flow rate in each case to quantify the desalination performance. The water flow rate is defined as the number of permeated water molecules divided by the time taken for permeation. In this work, the water flow rate was calculated using two methods which are described in Section S9, namely (a) using the slope of a linear fit of the $N_w$ versus time plot and (b) using the total number of permeated water molecules divided by the total time taken for their permeation. After comparison of calculated water flow rates obtained using the two methods in Section S9, we adopted method (a) to calculate the water flow rates in the rest of the article, because it better accounts for temporal fluctuations in the water flow rate. All the water flow rates and ion selectivities reported in this article were obtained as the average value from 5 independent simulations runs using different initial aqueous configurations and velocity distributions. The error bars indicated in each case represent the standard deviations from such collections of 5 runs.

To compare the calculated water flow rates with previously reported results, we plotted the obtained water flow rates as a function of the effective pressure drop and superimposed previously published results on the same plot, as illustrated in Figure 7. The effective pressure drop is given as ($P_{eff} = \Delta P - \Delta \Pi$), where $\Delta P$ and $\Delta \Pi$ represent, respectively, the difference in the applied pressures and the difference in the osmotic pressures, between the feed and permeate sides. Note that the $\Delta \Pi$ term has been neglected in previous simulation work on coupled water and ion transport through nanoporous hBN. This term is calculated using the Van't Hoff equation, $\Delta \Pi = iRT\Delta c$, where $i$ is the Van't Hoff factor (2 due to complete dissociation of NaCl), $\Delta c$ is the difference in the molar concentrations of the solute on the feed and permeate sides (610.863 mol/m$^3$ or 35.7 gm/L), $R$ is the universal gas constant (8.314 Pa m$^3$/mol K$^{-1}$), and $T$ is the temperature of the system (298.15 K), whence the calculated value of the osmotic pressure difference is 3.03 MPa. The three feed-side pressures, i.e., 50 MPa, 100 MPa, and 200 MPa lead to effective pressure values of 46.87 MPa, 96.87 MPa, and 196.87 MPa, respectively. Similar values of pressure drops have been considered in several theoretical studies.[17,18] Moreover, such high feed pressures (closer to 50 MPa) are currently also being explored experimentally.[78] Nevertheless, later in the article, we also report water permeabilities, which are normalized by the pressure drop, such that our results are independent of the applied pressure.

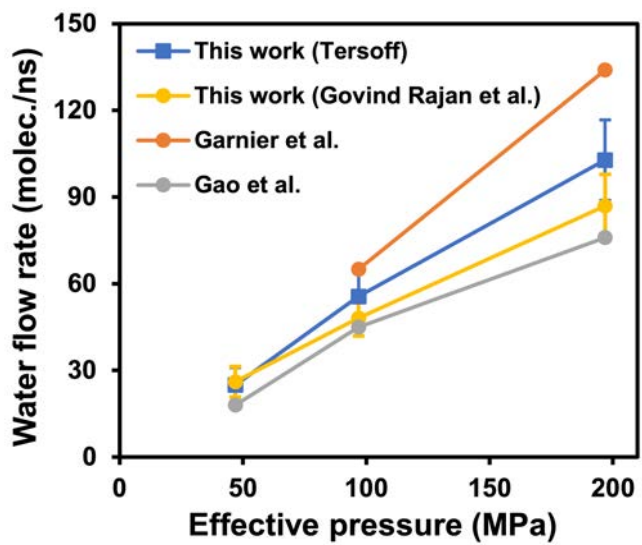

**Figure 7.** Water flow rate versus effective pressure plots for nanoporous monocrystalline hBN membrane. The calculations in this work were carried out using two different interatomic potentials to model hBN flexibility – the Tersoff force field proposed by Albe et al.,[45] with the results in purple color and the valence force field proposed by Govind Rajan et al.,[51] with the results in yellow color. Results from Garnier et al.[17] and Gao et al.[18] are shown in green and gray color, respectively. Lines are only visual guides and do not indicate any fit.

It can be inferred from Figure 7 that the water flow rates predicted in this work for nanoporous monocrystalline hBN are comparable with previously published work. Note that no previous calculations are available for water and ion permeation through nanoporous bicrystalline hBN. Nevertheless, variations were observed from previously predicted water flow rates of nanoporous monocrystalline hBN due to different force-field parameters employed, and the shape and size of the nanopores considered in previous work. For example, Gao et al.[18] considered the same nanopore used in this work but used a different water model (TIP3P), Lennard-Jones (LJ) parameters for water-hBN interactions, partial charges, and the force field to model the flexibility of hBN (universal force field). In contrast, Garnier et al.[17] used the TIP4P/2005 water model with the same LJ parameters as employed by Gao et al.[18], but considered an elongated nanopore. Since not only the size of the nanopore, but also its shape affects the permeation of water,[15] it is reasonable that the water flow rate predicted by Garnier et al.[17] is much higher than that predicted by Gao et al.[18] It can also be seen in Figure 7 that the force field used to model the flexibility of nanoporous monocrystalline hBN does not affect water permeation significantly, as seen from our results shown in purple (using the Tersoff force field proposed by Albe et al.[45]) and yellow color (using the valence force field proposed by Govind Rajan et al.[51]). After comparison of the water flow rates of nanoporous *monocrystalline* hBN membranes predicted by different force fields, the Tersoff force field was used to model the flexibility of nanoporous *monocrystalline* and *bicrystalline* hBN membranes in the rest of this work, because the force field by Govind Rajan et al.[51] cannot be used to model B-B and N-N bonds that are seen at GBs.

***Effect of GBs on the water flow rate, water flux, and water permeability through nanoporous hBN***. We next analyzed the effect of GBs on the desalination performance of nanoporous bicrystalline hBN membrane. As seen before, in Figure 6, the number of water molecules passing through nanoporous monocrystalline and bicrystalline hBN membranes were plotted against the simulation time at different feed pressures. Subsequently, we calculated the water flow rate for various bicrystalline hBN membrane configurations and plotted the values against the effective pressure, as shown in Figure 8a. One can infer from Figure 8a that the water flow rate follows an almost linear trend with the effective pressure; as the effective pressure increases, the water flow rate also increases for all the hBN membrane configurations. It is evident from Figure 8a that the lower-angle misoriented bicrystalline hBN membrane (with 13.2° misorientation angle) shows improved water flow rate as compared to monocrystalline hBN configuration, by ~30% at 46.87 MPa effective pressure. This finding can be attributed to the higher equivalent circular diameter (5.93 Å) and lower aspect ratio (0.84) of the nanopore in 13.2°-misoriented bicrystalline hBN, as compared to the nanopore of diameter 5.80 Å and aspect ratio 0.91 in monocrystalline hBN (see Table 1). On the other hand, the water flow rate of the higher-angle misoriented bicrystalline hBN membranes is comparable with that for the monocrystalline configuration. This observation is despite the fact that the nanopores in the higher-angle misoriented bicrystalline hBN membranes had a lower area (equivalent circular diameter) than the nanopore considered in monocrystalline hBN, thus indicating that GBs intrinsically enhance water permeation through nanoporous hBN. The underlying reason could be attributed to the lower aspect ratio of the nanopores in the 21.8°- and 32.2°-misoriented bicrystalline hBN, as compared to the nanopore in monocrystalline hBN, as seen in Table 1.

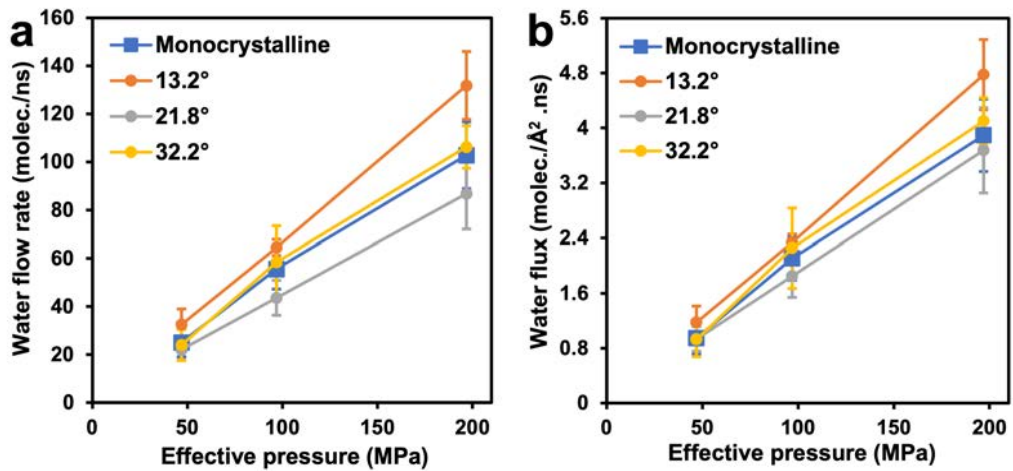

**Figure 8.** (a) Water flow rate and (b) water flux versus effective pressure for nanoporous monocrystalline and bicrystalline hBN membranes.

As mentioned above, the nanopore shape significantly affects water and ion permeation through nanoporous hBN. Thus, we maintained an approximately triangular shape of the nanopore in all the bicrystalline configurations, as in the monocrystalline configuration. Nevertheless, due to this constraint, the sizes of the nanopores slightly varied in the bicrystalline configurations as compared to the monocrystalline configuration. To analyze the intrinsic effect of GB configurations on the desalination performance of nanoporous bicrystalline hBN, the water *flux* (i.e., the water flow rate normalized by the nanopore area) was calculated and plotted as a function of the effective pressure in Figure 8b. It can be deduced from Figure 8b that the water flux follows a similar trend as followed by the water flow rate. Nevertheless, the improvement in the water flux was slightly smaller (~24%) at 46.87 MPa effective pressure, as compared to the improvement in the water flow rate, in the case of the 13.2°-misoriented bicrystalline hBN membrane. On the other hand, the water fluxes through the 21.8°- and 32.2°-misoriented bicrystalline hBN membranes were closer to the monocrystalline hBN case.

The water permeability in L cm$^{-2}$ day$^{-2}$ MPa$^{-1}$, defined as the volume of water permeated per unit time, membrane area, and pressure, was also calculated at 46.87 MPa effective pressure and plotted in Figure 9. It can be seen from Figure 9 that the improvement in the water permeability was more remarkable (~47%) at 46.87 MPa effective pressure, as compared to the improvement in the water flow rate, in the case of the 13.2°-misoriented bicrystalline hBN membrane. This extra improvement in the permeability of the 13.2°-misoriented bicrystalline hBN membrane is attributed to its lower surface area (1000 Å$^2$), as compared to the higher surface area of the monocrystalline hBN membrane (1136 Å$^2$). On the other hand, the water permeabilities through the 21.8°- and 32.2°-misoriented bicrystalline hBN membranes are almost identical to that of the monocrystalline hBN membrane. Nevertheless, the water permeability estimate of ~12-18 L cm$^{-2}$ day$^{-2}$ MPa$^{-1}$ for nanoporous hBN is comparable in order of magnitude to estimates obtained for graphene,[6] MoS$_2$,[23] and MOF[22] membranes considering similar pore areas as in our study (25-30 Å$^2$, see Table 1).

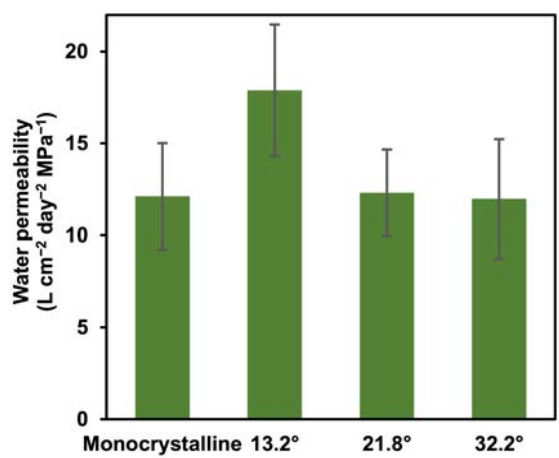

**Figure 9.** Water permeability at 46.87 MPa effective pressure for monocrystalline and bicrystalline hBN membranes.

***Comparison of our results with experimental data and avenues for future work.*** The calculated values of the water permeability for monocrystalline and bicrystalline hBN membranes are in the range of 12.0 to 17.9 L cm$^{-2}$ day$^{-2}$ MPa$^{-1}$, which are several times higher than those of current commercial RO membranes (~0.1 L cm$^{-2}$ day$^{-1}$ MPa$^{-1}$, see ref. 6) and better than that of a graphene filter (~5.9 L cm$^{-2}$ day$^{-2}$ MPa$^{-1}$ measured in experiment[9]). However, as mentioned before, the pore density considered in our work is ~10$^{17}$ nm$^{-2}$. This value, although comparable to what is considered in other simulation studies, is about an order of magnitude higher than the value of 10$^{16}$ nm$^{-2}$ achieved experimentally in graphene.[9] This observation indicates that one may reasonably achieve anywhere between 1.2 to 1.8 L cm$^{-2}$ day$^{-2}$ MPa$^{-1}$ of performance using a hBN membrane. (A better performance would require the generation of more pores per unit area in hBN.) In contrast, the experimentally measured values of water permeance for amino-functionalized nanoporous hBN are in the range of 14.9 to ~48 L cm$^{-2}$ day$^{-2}$ MPa$^{-1}$ for membranes of thickness 2000 nm and 400 nm, respectively.[12] The variation in the water permeability by a factor of 2-3 between the simulated and experimental results can be attributed to the larger sizes of the nanopores in the experimental samples. Indeed, the sizes of the nanopores in the experimentally synthesized hBN used by Chen et al.[12] are in the range of 8.0 to 18.3 Å, which are larger than the sizes of the nanopores considered in our work (5 to 6 Å). Note that the study of Chen et al.[12] considered lamellar hBN membranes, i.e., membranes synthesized using exfoliated hBN flakes. In addition, work is needed to understand how edge functionalization affects the sizes of nanopores in hBN, and consequently water and ion transport through nanopores in the material. The investigation of edge functional groups would also require the development of force-field parameters to describe functionalized hBN edges.

***Quantifying the free energy barriers for water permeation.*** When water molecules pass through a nanopore in the hBN membrane, they have to surmount a free energy barrier. To further understand the water permeation mechanism through nanopores, we calculated the free energy profile or potential of mean force (PMF) of water molecules using the Boltzmann sampling method.[18,26,79] According to the Boltzmann sampling method, the free energy or PMF is given as, $F(z) = -RT\ln\rho(z)$, where $R$ represents the universal gas constant, $T$ represents the temperature, and $\rho(z)$ represents the density at position $z$ in the system. After obtaining the water density profiles from an equilibrium simulation with equal feed and permeate pressures of 1 bar, the free energy profiles were plotted for nanoporous monocrystalline and bicrystalline hBN membranes in Figure 10.

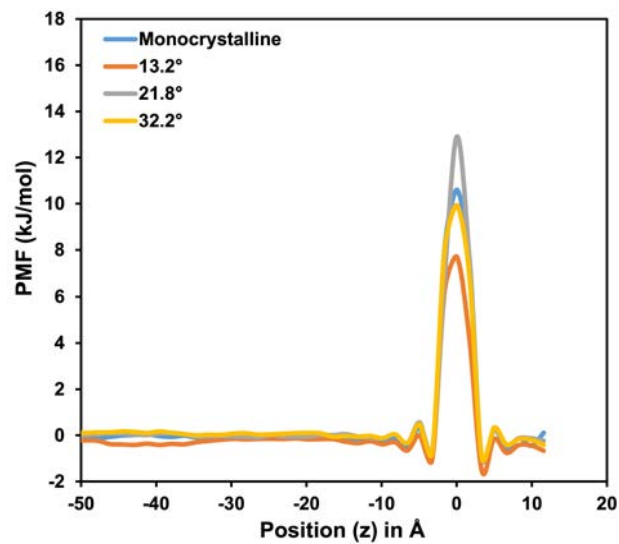

**Figure 10.** Free energy profile or PMF variations along the flow direction (*z*-direction) for nanoporous monocrystalline and bicrystalline hBN membranes of varying misorientation angles.

It can be inferred from Figure 10 that the free energy profile of the water molecules is approximately constant far away from the membrane, which is placed at $z = 0$, in all cases. At the position of membrane ($z = 0$), a large increase in the free energy is observed. This free energy enhancement at the nanopore represents the energy barrier for water permeation from the feed side to the permeate side. It can be observed from Figure 10 that the lower-angle misoriented bicrystalline hBN membrane shows a lower energy barrier, which causes higher water flux, as compared to the higher-angle misoriented bicrystalline configuration. In contrast, the 21.8°-misoriented bicrystalline hBN membrane shows a higher energy barrier, such that the water flux is lower. Overall, the trends presented by the free energy barriers (21.8° > 0° > 32.2° > 13.2°) in Figure 10 match the trends presented by the water fluxes (21.8° < 0° < 32.2° < 13.2°), where a misorientation of 0° indicates nanoporous monocrystalline hBN. Moreover, these results support the observation that, even after accounting for differences in nanopore areas, the 13.2°-misoriented bicrystalline hBN allows increased water flux through the hBN membrane. In this regard, future work could investigate the coupled role of temperature and grain boundaries on modulating water and ion transport through 2D nanopores, by simulating the desalination process at various system temperatures. Such an approach would enable the estimation of activation barriers for water permeation using an alternative method and the ensuing results could be compared with the PMF-based approach presented here.

***Effect of GBs on ion rejection by nanoporous hBN.*** Apart from the water flux, the extent of rejection of $Na^+$ and $Cl^-$ ions from water is a crucial criterion for any desalination membrane. We counted the percentage of $Na^+$ and $Cl^-$ ions that were rejected in each simulation run for various membrane configurations at different effective pressure values, as summarized in Tables 2 and 3, respectively. It can be deduced from Table 2 that, in only the monocrystalline membrane configuration, the sodium rejection is greater than 99% (corresponding to a maximum of a single $Na^+$ ion permeating in one out of five independent simulation runs). For the bicrystalline membranes, the $Na^+$ rejection follows the order: 32.2° > 21.8° > 13.2°, with the maximum possible rejection being 97.3%. The 13.2°-misoriented bicrystalline membrane shows the least $Na^+$ rejection as compared to monocrystalline hBN configuration, which is only 92.7% at 46.87 MPa effective pressure. Nevertheless, we observe from Table 3 that as such, no $Cl^-$ ions (with an ionic radius of 1.81 Å[80]) pass through the membrane (except in very few runs) due to their larger size (note that the ionic radius of $Na^+$ is 0.99 Å[80]). Since a high ion rejection (>99%) is an important requirement, and is considered even more significant than high water permeability for commercial RO membranes,[81] our findings indicate that the presence of GBs in nanoporous hBN can compromise their water desalination performance through a drop in ion rejection.

Table 2. Percentage of $Na^+$ ions rejected through the different membrane configurations at various effective pressures.

| Pressure (MPa) | $Na^+$ ions rejection (%) | | | |
|---|---|---|---|---|
| | Monocrystalline nanoporous hBN | Bicrystalline nanoporous hBN | | |
| | | 13.2° | 21.8° | 32.2° |
| 46.87 | 99.09 ± 2.03 | 92.73 ± 6.90 | 95.45 ± 0.00 | 97.27 ± 2.49 |
| 96.87 | 99.09 ± 2.03 | 92.73 ± 5.18 | 95.45 ± 0.00 | 97.27 ± 2.49 |
| 196.87 | 97.27 ± 2.49 | 92.72 ± 4.06 | 95.45 ± 0.00 | 95.45 ± 0.00 |

Table 3. Percentage of $Cl^-$ ions rejected through the different membrane configurations at various effective pressures.

| Pressure (MPa) | $Cl^-$ ions rejection (%) | | | |
|---|---|---|---|---|
| | Monocrystalline nanoporous hBN | Bicrystalline nanoporous hBN | | |
| | | 13.2° | 21.8° | 32.2° |
| 46.87 | 100 ± 0.00 | 99.09 ± 2.03 | 100 ± 0.00 | 100 ± 0.00 |
| 96.87 | 100 ± 0.00 | 100 ± 0.00 | 100 ± 0.00 | 99.09 ± 2.03 |
| 196.87 | 99.09 ± 2.03 | 100 ± 0.00 | 100 ± 0.00 | 100 ± 0.00 |

***Effect of interfacial electrostatic interactions on water desalination performance.*** Although, intuitively, interfacial electrostatics should play an important role in modulating water and ion permeation, no previous study has investigated the effect of these electrostatic interactions on the desalination performance of through nanoporous hBN. In this regard, the method used to assign partial charges to the membrane atoms, based on DFT calculations, is crucial. In previous work, authors have either used Mulliken population analysis[82] (e.g., Davoy et al.[15] and Garnier et al.[17]) or the Hirshfeld charge scheme[83] (e.g., Gao et al.[18]) to obtain the charges on B and N atoms. Yet others have used the density derived electrostatic and chemical (DDEC) method[84] and electrostatic-potential-fitting (ESP) method (e.g., Jafarzadeh et al.[10]) to calculate partial charges. In previous work, we showed that ESP-based charges are not able to predict correctly the electrostatic potential above defects in hBN.[85] In Section S10, we show that DDAP (using default parameters), Hirshfeld, Mulliken, and DDEC partial charges are also not suitable for predicting the electrostatic potential above nanopores in hBN. Instead, only the DDAP charges with customized parameters are found to be accurate for this purpose. Finally, Loh[4] investigated the desalination performance of nanoporous hBN membrane by using the partial charges on B and N atoms from periodic BN nanotubes, which as we show below, may underpredict the water permeation rate through nanoporous hBN. To this end, here, we analyzed the effect of electrostatic interactions on the desalination performance of monocrystalline and bicrystalline nanoporous hBN. To evaluate the effect of interfacial electrostatic interactions, we calculated the water flow rate with three different sets of partial charges on the B and N atoms in hBN membrane: (i) partial charges calculated using DFT and the DDAP method, (ii) bulk partial charges on each B ($q = +0.907\ e$) and N atom ($q = -0.907\ e$) (these charges were calculated by Govind Rajan et al.[51] for monocrystalline hBN nanosheets), and (iii) by taking zero charges on each B and N atom, i.e., when the interfacial electrostatic interactions are zero. After calculating the water flow rate for different sets of partial charges, we plotted these results against the effective pressure for various nanoporous monocrystalline and bicrystalline hBN membranes in Figure 11. The results for the water flow rate are also tabulated in Table 4 for the different sets of partial charges considered at 196.87 MPa effective pressure. It can be inferred from Figure 11 and Table 4 that in all the nanoporous hBN membranes, the water flow rate is significantly affected by the interfacial electrostatic interactions. The values of water flow rate were highest when the altered partial charges on B and N atoms were calculated using DFT-based simulations and the DDAP method, whereas the values of water flow rate were lowest when bulk charges (charges on monocrystalline hBN nanosheets) on B and N atoms were used. (Note that, in the latter case, the hBN membrane is charged, due to the presence of an unequal number of B and N atoms in it. Accordingly, MD simulation packages add a uniform background charge to the system such that the net charge on the simulation box is zero.) The water flow rates calculated without considering the electrostatic interactions (i.e., considering zero partial charges on B and N atoms) were found to lie in between cases (i) and (ii). We also see in Table 4 that a maximum of 69.1% improvement in water flow rate was observed when compared with the flow rate predicted using zero partial charges and 297.4% improvement was observed when compared with the flow rate predicted using bulk charges. Overall, we conclude that the calculated partial charges must be accurate enough, as the water flow rate is significantly affected by the interfacial electrostatic interactions. Moreover, the water flow rate reduces significantly if the hBN sheet is charged. Note that, in our simulations, the monocrystalline, 13.2°-misoriented bicrystalline, and 21.8°-misoriented bicrystalline hBN configurations have 4 fewer B atoms than N atoms, leading to a surface charge density of -0.051 to -0.059 C m$^{-2}$. On the other hand, the 32.2°-misoriented hBN membrane has 5 fewer B atoms than N atoms, resulting in -0.066 C m$^{-2}$ of surface charge. It can also be seen from Table 4 that the water flow rate is maximum in 13.2°-misoriented bicrystalline membrane as compared to other membrane configurations even when we are considering the zero or bulk charges. Therefore, the water flow rate is always maximum in 13.2°-misoriented bicrystalline membrane, irrespective of whether we consider DFT, bulk, or zero charges for the B and N atoms. This finding indicates that our conclusion regarding the water permeability being highest through the 13.2°-misoriented bicrystalline hBN membrane is not affected by the minor difference between the DFT and partial-charge-based potential as seen in Figure 5b.

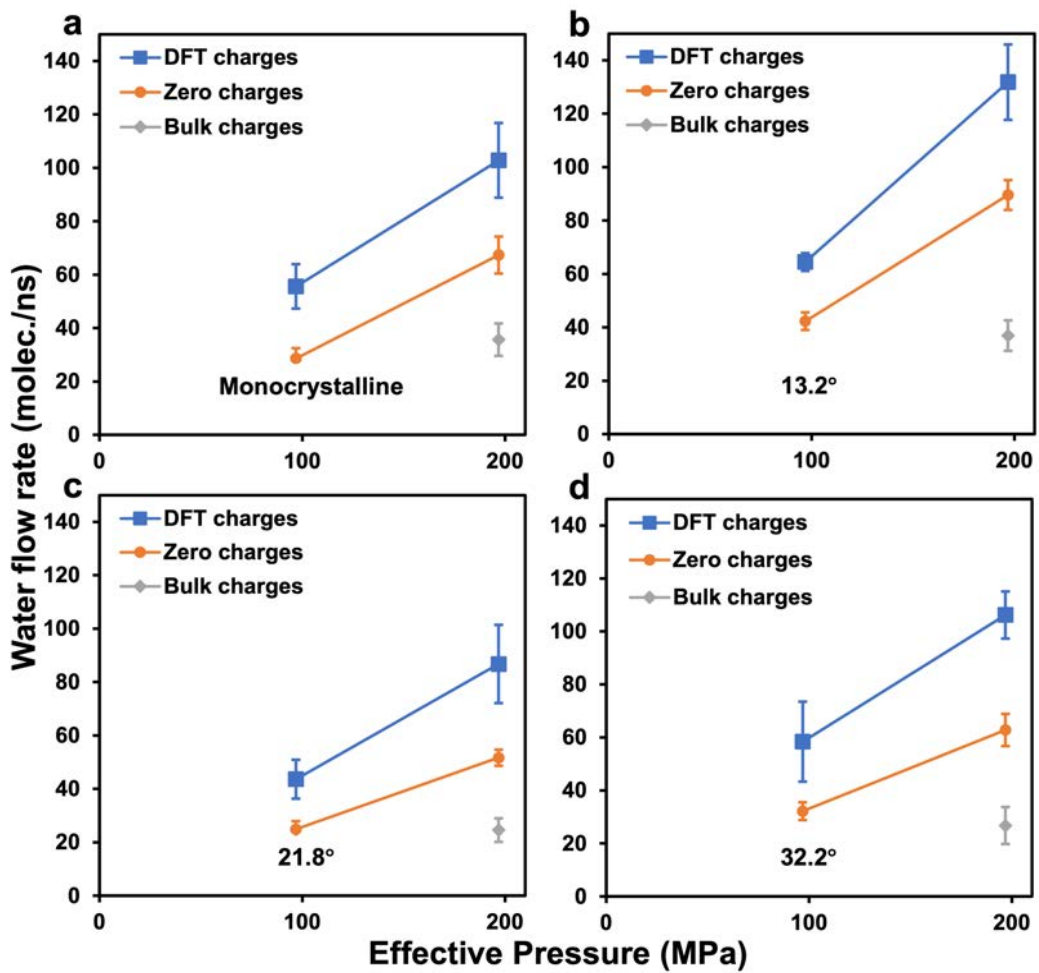

**Figure 11.** Water flow rate versus effective pressure plots for various configurations: (a) monocrystalline, (b) 13.2°-misoriented bicrystalline, (c) 21.8°-misoriented bicrystalline, and (d) 32.2°-misoriented bicrystalline nanoporous hBN membranes.

**Table 4.** Water flow rate through monocrystalline and bicrystalline hBN membranes for different partial charges at 196.87 MPa effective pressure.

| Configuration | Water flow rate (molec./ns) | | | % Improvement in water flow rate using DDAP charges | |
|---|---|---|---|---|---|
| | DFT charges | Zero charges | Bulk charges | Compared with zero charges | Compared with bulk charges |
| **Monocrystalline** | 102.8 | 67.4 | 35.7 | 52.6 | 188.3 |
| **13.2°** | 131.8 | 89.6 | 36.9 | 47.2 | 257.4 |
| **21.8°** | 86.7 | 51.7 | 24.6 | 67.9 | 253.3 |
| **32.2°** | 106.2 | 62.8 | 26.7 | 69.1 | 297.4 |

To further investigate the physics behind the effect of interfacial electrostatic interactions, we plotted the contours of interfacial interactions for nanoporous monocrystalline and 32.2°-misoriented bicrystalline hBN membranes using different sets of partial charges, as shown in Figure 12 and 13. To plot the contours of interfacial interactions between nanoporous hBN membrane and water molecule, initially, a water molecule was placed at the corner of the nanoporous hBN membrane at a distance of 3 Å above the surface, as depicted in Section S5. Subsequently, the water molecule was moved in the lateral direction and the interaction energy was calculated in a grid above the hBN surface. The interfacial interactions contours for 13.2°- and 21.8°-misoriented bicrystalline membranes are not depicted, as they show similar trends, as observed for the 32.2°-misoriented bicrystalline membrane. It can be seen from Figure 12 and 13 that the partial charges cause positive and negative patches of electrostatic potential in the vicinity of the nanopore. Depending on the magnitude of these patches, water permeation is enhanced (using DDAP-based DFT partial charges) or impeded (using bulk partial charges). Comparing panels b-c in Figure 12 with panels d-e in the same figure, we see that the water-nanoporous monocrystalline interaction energy is lower while using DDAP-based DFT partial charges (Figure 12b-c), as compared to while using bulk partial charges (Figure 12d-e). A similar conclusion is reached for the case of 32.2°-misoriented bicrystalline nanoporous hBN, by examining panels b-c in Figure 13 and comparing them with panels d-e in the same figure. Thus, the presence of an optimal level of interfacial Coulombic interactions enhances water transport to the surface, and through to the other side of the nanopore, as in the case with DDAP-based partial charges. However, when the interfacial Coulombic interactions are too high, as in the case with bulk partial charges, the water molecules get trapped at the nanopore mouth, leading to a reduced water permeation rate.

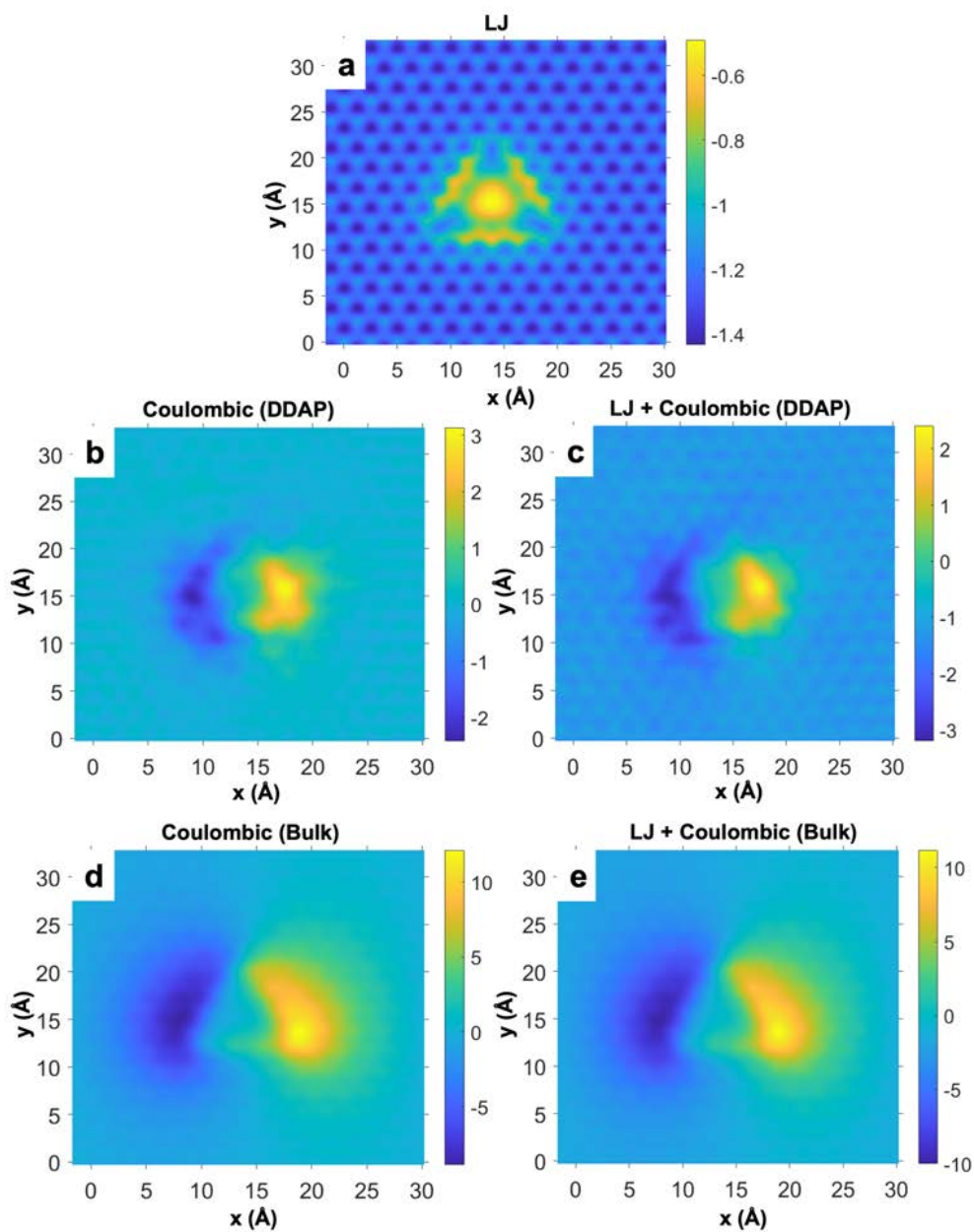

**Figure 12.** Interfacial interaction energy contours (in kcal/mol) for a tangential water molecule at 3 Å from the membrane, using different sets of partial charges for a nanoporous monocrystalline hBN membrane: (a) LJ interaction energy, (b) electrostatic potential energy calculated using DDAP-based DFT charges, (c) total potential energy calculated using DDAP-based DFT charges, (d) electrostatic potential energy calculated using bulk charges, and (e) total potential energy calculated using bulk charges.

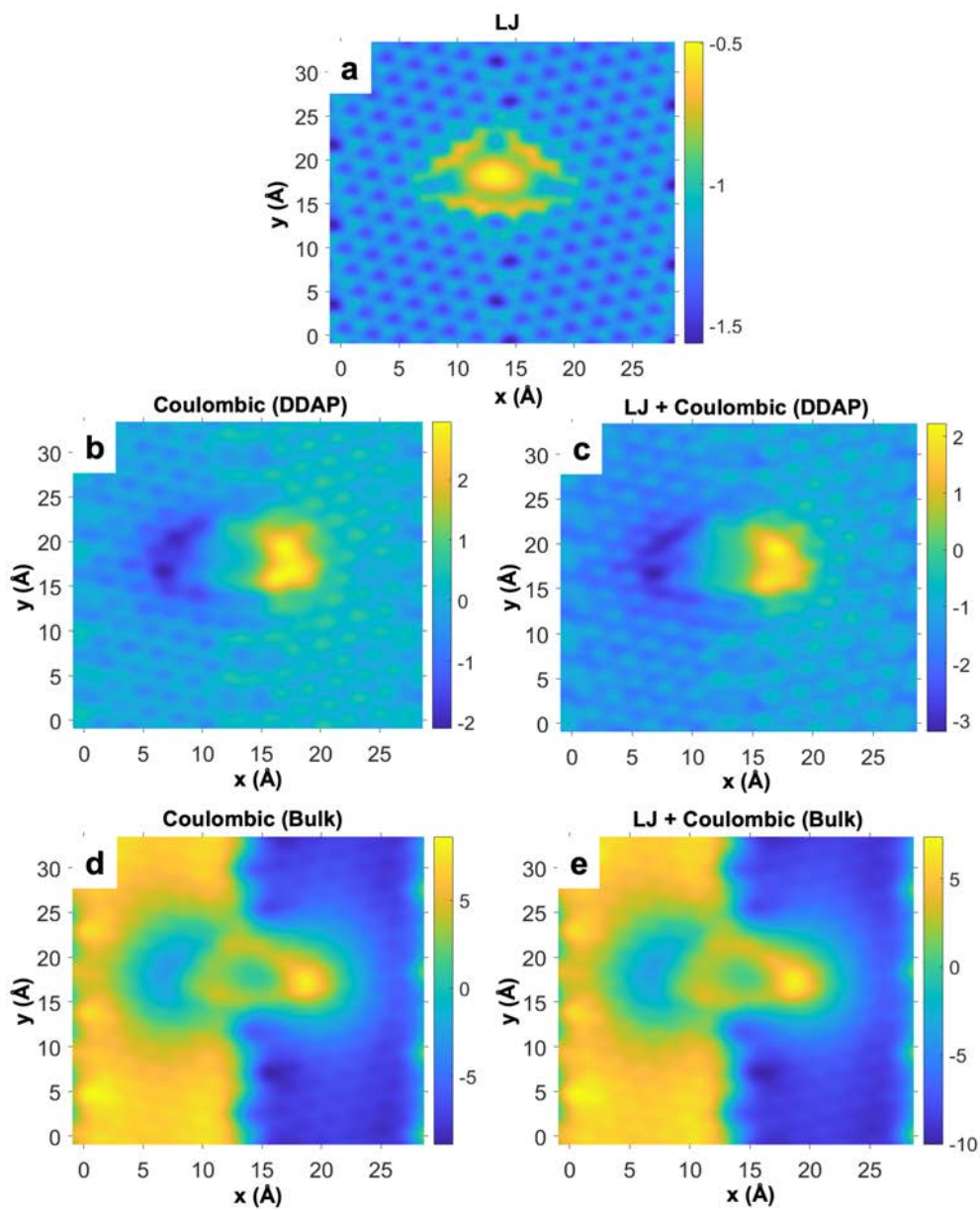

**Figure 13.** Interfacial interaction energy contours (in kcal/mol) for a tangential water molecule at 3 Å from the membrane, using different sets of partial charges for a 32.2°-misoriented bicrystalline nanoporous hBN membrane: (a) LJ interaction energy, (b) electrostatic potential energy calculated using DDAP-based DFT charges, (c) total potential energy calculated using DDAP-based DFT charges, (d) electrostatic potential energy calculated using bulk charges, and (e) total potential energy calculated using bulk charges.

**CONCLUSIONS**

For water desalination membranes, large-area 2D materials are required, which inevitably will contain grain boundaries (GBs) as geometrical defects. In this work, we investigated the role of GBs and interfacial electrostatic interactions in modulating the desalination performance of bicrystalline hBN membranes using classical MD simulations supported by quantum-mechanical DFT calculations. Before investigating the desalination performance, we calculated the partial charges on each B and N atom in hBN using DFT and validated these partial charges with the help of electrostatic potential plots. The electrostatic potential calculated using DFT and that using partial charges are in good agreement for monocrystalline and bicrystalline hBN configurations. By investigating water and ion permeation using MD simulations, we concluded that, although the lower-angle (13.2°) misoriented bicrystalline hBN membrane showed improved water flow rate by ~30%, the $Na^+$ rejection was reduced by ~6% as compared to monocrystalline hBN. The finding regarding increased water flow persisted even after correcting the water flow rates for the nanopore area, i.e., considering the water flux through the investigated nanopores. In fact, the water flow rate and flux through higher-angle (21.8° and 32.2°) misoriented bicrystalline hBN was found comparable to that through nanoporous monocrystalline hBN, despite them having smaller nanopore areas. We showed that these nanopores have a lower aspect ratio, thus pointing to the role of the nanopore shape in modulating water flux, as well as ion rejection. Indeed, the increased water flow rate/flux was accompanied by a significant drop in $Na^+$ rejection by the membrane. We also deduced that the values of the water flow rate were highest when altered partial charges on B and N atoms were calculated using DDAP-based DFT calculations, whereas the values of water flow rate were lowest when bulk partial charges (i.e., the charges in a monocrystalline hBN nanosheet) on B and N atoms were used. This implies that surface charge on the membrane reduces the rate of water permeation. Additionally, the water flow rates calculated without considering interfacial electrostatic interactions, i.e., by taking zero partial charges on B and N atoms were in between the two cases discussed above. Overall, our investigation of the role of GBs and interfacial electrostatic interactions in modulating the desalination performance of nanoporous bicrystalline hBN informs the use of large-area hBN membrane in seawater desalination applications. Specifically, we find that the presence of GBs and surface charge may be deleterious for, respectively, ion rejection and water permeation through nanoporous hBN membranes. Thus, monocrystalline nanoporous hBN should be preferred for synthesizing water desalination membranes. To conclude, we hope that future studies consider the effect of GBs, which are prevalent in large-area membranes, on the desalination performance of other nanoporous 2D materials as well, while realistically modeling interfacial electrostatic interactions.

**SUPPORTING INFORMATION**

Water flow rate for nanoporous monocrystalline hBN membrane at 50 MPa feed pressure when the permeate side has 500 and 2000 water molecules; convergence of energy and temperature during equilibration; parameters for the LJ and Coulombic interatomic potentials; choice of the force fields for hBN, water molecules, salt ions, and interfacial interactions between hBN and water; model used to calculate the interfacial interactions between a nanoporous hBN membrane and a unit test charge/tangential water molecule; nomenclature and repeat length of the grain boundaries; calculation of the areas, equivalent circular diameters, and aspect ratios of the various nanopores considered; tuning of partial charges and validation of the resultant electrostatic potential for 13.2°-misoriented bicrystalline nanoporous hBN; comparing two different methods for water flow rate calculation; and comparison of the classical electrostatic potentials calculated using DDAP, Hirshfeld, Mulliken, and density derived electrostatic and chemical (DDEC) partial charges with the DFT-derived potential.

**CONFLICTS OF INTEREST**

There are no conflicts of interest to declare.

ACKNOWLEDGEMENTS

We gratefully acknowledge the financial support received from the National Supercomputing Mission (DST/NSM/R&D_HPC_Applications/2021/07), which is coordinated by the Department of Science and Technology (DST) and the Department of Electronics and Information Technology (DeitY). We also thank the Supercomputer Education and Research Centre (SERC) at the Indian Institute of Science for computational support. A.G.R. thanks Prof. E. D. Jemmis, Prof. Sudeep Punnathanam, and Dr. Rahul Prasanna Misra for discussions on charge separation between B and N atoms at grain boundaries in hBN, as well as Prof. Yagnaseni Roy for information on the ion rejection performance of commercial RO membranes.

**REFERENCES**

(1) Elimelech, M.; Phillip, W. A. The Future of Seawater Desalination: Energy, Technology, and the Environment. *Science* **2011**, *333*, 712–717.

(2) Shannon, M. A.; Bohn, P. W.; Elimelech, M.; Georgiadis, J. G.; Mariñas, B. J.; Mayes, A. M. Science and Technology for Water Purification in the Coming Decades. *Nature* **2008**, *452*, 301–310.

(3) Xu, G.-R.; Xu, J.-M.; Su, H.-C.; Liu, X.-Y.; Lu-Li; Zhao, H.-L.; Feng, H.-J.; Das, R. Two-Dimensional (2D) Nanoporous Membranes with Sub-Nanopores in Reverse Osmosis Desalination: Latest Developments and Future Directions. *Desalination* **2019**, *451*, 18–34.

(4) Loh, G. C. Fast Water Desalination by Carbon-Doped Boron Nitride Monolayer: Transport Assisted by Water Clustering at Pores. *Nanotechnology* **2019**, *30*, 055401.

(5) Water Scarcity: Overview. *World Wildlife Fund (WWF) Inc.* https://worldwildlife.org/threats/water-scarcity. (Last accessed: February 2022).

(6) Cohen-Tanugi, D.; Grossman, J. C. Water Desalination across Nanoporous Graphene. *Nano Lett.* **2012**, *12*, 3602–3608.

(7) Homaeigohar, S.; Elbahri, M. Graphene Membranes for Water Desalination. *NPG Asia Mater.* **2017**, *9*, e427–e427.

(8) Cohen-Tanugi, D.; Grossman, J. C. Nanoporous Graphene as a Reverse Osmosis Membrane: Recent Insights from Theory and Simulation. *Desalination* **2015**, *366*, 59–70.

(9) Surwade, S. P.; Smirnov, S. N.; Vlassiouk, I. V.; Unocic, R. R.; Veith, G. M.; Dai, S.; Mahurin, S. M. Water Desalination Using Nanoporous Single-Layer Graphene. *Nat. Nanotechnol.* **2015**, *10*, 459–464.

(10) Jafarzadeh, R.; Azamat, J.; Erfan-Niya, H.; Hosseini, M. Molecular Insights into Effective Water Desalination through Functionalized Nanoporous Boron Nitride Nanosheet Membranes. *Appl. Surf. Sci.* **2019**, *471*, 921–928.

(11) Teow, Y. H.; Mohammad, A. W. New Generation Nanomaterials for Water Desalination: A Review. *Desalination* **2019**, *451*, 2–17.

(12) Chen, C.; Wang, J.; Liu, D.; Yang, C.; Liu, Y.; Ruoff, R. S.; Lei, W. Functionalized Boron Nitride Membranes with Ultrafast Solvent Transport Performance for Molecular Separation. *Nat. Commun.* **2018**, *9*, 1902.

(13) Köhler, M. H.; Bordin, J. R.; Barbosa, M. C. 2D Nanoporous Membrane for Cation Removal from Water: Effects of Ionic Valence, Membrane Hydrophobicity, and Pore Size. *J. Chem. Phys.* **2018**, *148*, 222804.

(14) Prasad K., V. P.; Kannam, S. K.; Hartkamp, R.; Sathian, S. P. Water Desalination Using Graphene Nanopores: Influence of the Water Models Used in Simulations. *Phys. Chem. Chem. Phys.* **2018**, *20*, 16005–16011.

(15) Davoy, X.; Gellé, A.; Lebreton, J.-C.; Tabuteau, H.; Soldera, A.; Szymczyk, A.; Ghoufi, A. High Water Flux with Ions Sieving in a Desalination 2D Sub-Nanoporous Boron Nitride Material. *ACS Omega* **2018**, *3*, 6305–6310.

(16) Sharma, B. B.; Parashar, A. A Review on Thermo-Mechanical Properties of Bi-Crystalline and Polycrystalline 2D Nanomaterials. *Crit. Rev. Solid State Mater. Sci.* **2020**, *45*, 134–170.

(17) Garnier, L.; Szymczyk, A.; Malfreyt, P.; Ghoufi, A. Physics behind Water Transport through Nanoporous Boron Nitride and Graphene. *J. Phys. Chem. Lett.* **2016**, *7*, 3371–3376.

(18) Gao, H.; Shi, Q.; Rao, D.; Zhang, Y.; Su, J.; Liu, Y.; Wang, Y.; Deng, K.; Lu, R. Rational Design and Strain Engineering of Nanoporous Boron Nitride Nanosheet Membranes for Water Desalination. *J. Phys. Chem. C* **2017**, *121*, 22105–22113.

(19) Sharma, B. B.; Parashar, A. Mechanical Strength of a Nanoporous Bicrystalline H-BN Nanomembrane in a Water Submerged State. *Phys. Chem. Chem. Phys.* **2020**, *22*, 20453–20465.

(20) Faucher, S.; Aluru, N.; Bazant, M. Z.; Blankschtein, D.; Brozena, A. H.; Cumings, J.; Pedro de Souza, J.; Elimelech, M.; Epsztein, R.; Fourkas, J. T.; Rajan, A. G.; Kulik, H. J.; Levy, A.; Majumdar, A.; Martin, C.; McEldrew, M.; Misra, R. P.; Noy, A.; Pham, T. A.; Reed, M.; Schwegler, E.; Siwy, Z.; Wang, Y.; Strano, M. Critical Knowledge Gaps in Mass Transport through Single-Digit Nanopores: A Review and Perspective. *J. Phys. Chem. C* **2019**, *123*, 21309–21326.

(21) Wang, Y.; He, Z.; Gupta, K. M.; Shi, Q.; Lu, R. Molecular Dynamics Study on Water Desalination through Functionalized Nanoporous Graphene. *Carbon N. Y.* **2017**, *116*,

120–127.

(22) Cao, Z.; Liu, V.; Farimani, A. B. Water Desalination with Two-Dimensional Metal-Organic Framework Membranes. *Nano Lett.* **2019**, *19*, 8638–8643.

(23) Heiranian, M.; Farimani, A. B.; Aluru, N. R. Water Desalination with a Single-Layer MoS2 Nanopore. *Nat. Commun.* **2015**, *6*, 8616.

(24) Konatham, D.; Yu, J.; Ho, T. A.; Striolo, A. Simulation Insights for Graphene-Based Water Desalination Membranes. *Langmuir* **2013**, *29*, 11884–11897.

(25) Cohen-Tanugi, D.; Grossman, J. C. Water Permeability of Nanoporous Graphene at Realistic Pressures for Reverse Osmosis Desalination. *J. Chem. Phys.* **2014**, *141*, 074704.

(26) Cohen-tanugi, D.; Lin, L.; Grossman, C. Multilayer Nanoporous Graphene Membranes for Water Desalination. **2016**.

(27) Yang, X.; Yang, X.; Liu, S. Molecular Dynamics Simulation of Water Transport through Graphene-Based Nanopores: Flow Behavior and Structure Characteristics. *Chinese J. Chem. Eng.* **2015**, *23*, 1587–1592.

(28) Li, Y.; Xu, Z.; Liu, S.; Zhang, J.; Yang, X. Molecular Simulation of Reverse Osmosis for Heavy Metal Ions Using Functionalized Nanoporous Graphenes. *Comput. Mater. Sci.* **2017**, *139*, 65–74.

(29) Vishnu Prasad, K.; Sathian, S. P. The Effect of Temperature on Water Desalination through Two-Dimensional Nanopores. *J. Chem. Phys.* **2020**, *152*, 164701.

(30) Liu, Y.; Zhao, Y.; Zhang, X.; Huang, X.; Liao, W.; Zhao, Y. $MoS_2$-Based Membranes in Water Treatment and Purification. *Chem. Eng. J.* **2021**, *422*, 130082.

(31) Li, H.; Zeng, X. C. Wetting and Interfacial Properties of Water Nanodroplets in Contact with Graphene and Monolayer Boron Nitride Sheets. *ACS Nano* **2012**, *6*, 2401–2409.

(32) Wu, Y.; Wagner, L. K.; Aluru, N. R. Hexagonal Boron Nitride and Water Interaction Parameters. *J. Chem. Phys.* **2016**, *144*, 164118.

(33) Tsukanov, A. A.; Shilko, E. V. Computer-Aided Design of Boron Nitride-Based Membranes with Armchair and Zigzag Nanopores for Efficient Water Desalination. *Materials (Basel).* **2020**, *13*, 1–12.

(34) Liu, L.; Liu, Y.; Qi, Y.; Song, M.; Jiang, L.; Fu, G.; Li, J. Hexagonal Boron Nitride with Nanoslits as a Membrane for Water Desalination: A Molecular Dynamics Investigation. *Sep. Purif. Technol.* **2020**, *251*, 117409.

(35) Azamat, J.; Sardroodi, J. J.; Poursoltani, L.; Jahanshahi, D. Functionalized Boron Nitride Nanosheet as a Membrane for Removal of $Pb^{2+}$ and $Cd^{2+}$ Ions from Aqueous Solution. *J. Mol. Liq.* **2021**, *321*, 114920.

(36) Majidi, S.; Pakdel, S.; Azamat, J.; Erfan-Niya, H. Hexagonal Boron Nitride (h-BN) in Solutes Separation. In; Das, R., Ed.; Springer International Publishing: Cham, 2021; pp.

163–191.

(37) Azamat, J.; Khataee, A.; Joo, S. W. Separation of Copper and Mercury as Heavy Metals from Aqueous Solution Using Functionalized Boron Nitride Nanosheets: A Theoretical Study. *J. Mol. Struct.* **2016**, *1108*, 144–149.

(38) Lei, W.; Portehault, D.; Liu, D.; Qin, S.; Chen, Y. Porous Boron Nitride Nanosheets for Effective Water Cleaning. *Nat. Commun.* **2013**, *4*, 1777.

(39) Srivastava, R.; Kommu, A.; Sinha, N.; Singh, J. K. Removal of Arsenic Ions Using Hexagonal Boron Nitride and Graphene Nanosheets: A Molecular Dynamics Study. *Mol. Simul.* **2017**, *43*, 985–996.

(40) Sharma, B. B.; Parashar, A. Atomistic Simulations to Study the Effect of Water Molecules on the Mechanical Behavior of Functionalized and Non-Functionalized Boron Nitride Nanosheets. *Comput. Mater. Sci.* **2019**, *169*, 109092.

(41) Plimpton, S. Fast Parallel Algorithms for Short-Range Molecular Dynamics. *J. Comput. Phys.* **1995**, *117*, 1–19.

(42) Nosé, S. A Unified Formulation of the Constant Temperature Molecular Dynamics Methods. *J. Chem. Phys.* **1984**, *81*, 511–519.

(43) Hoover, W. G. Canonical Dynamics: Equilibrium Phase-Space Distributions. *Phys. Rev. A* **1985**, *31*, 1695–1697.

(44) Hockney, R. .; Eastwood, J. . Particle-Particle–Particle-Mesh (P3M) Algorithms. In *Computer Simulation Using Particles*; CRC Press, 1988; pp. 267–304.

(45) Albe, K.; Möller, W.; Heinig, K. Computer Simulation and Boron Nitride. *Radiat. Eff. Defects Solids* **1997**, *141*, 85–97.

(46) Abascal, J. L. F.; Vega, C. A General Purpose Model for the Condensed Phases of Water: TIP4P/2005. *J. Chem. Phys.* **2005**, *123*, 234505.

(47) Li, Y.; Wei, A.; Ye, H.; Yao, H. Mechanical and Thermal Properties of Grain Boundary in a Planar Heterostructure of Graphene and Hexagonal Boron Nitride. *Nanoscale* **2018**, *10*, 3497–3508.

(48) Sharma, B. B.; Parashar, A. Inter-Granular Fracture Behaviour in Bicrystalline Boron Nitride Nanosheets Using Atomistic and Continuum Mechanics-Based Approaches. *J. Mater. Sci.* **2021**, *56*, 6235–6250.

(49) Ding, Q.; Ding, N.; Liu, L.; Li, N.; Wu, C.-M. L. Investigation on Mechanical Performances of Grain Boundaries in Hexagonal Boron Nitride Sheets. *Int. J. Mech. Sci.* **2018**, *149*, 262–272.

(50) Jones, J. E. On the Determination of Molecular Fields. —II. From the Equation of State of a Gas. *Proc. R. Soc. London. Ser. A, Contain. Pap. a Math. Phys. Character* **1924**, *106*, 463–477.

(51) Govind Rajan, A.; Strano, M. S.; Blankschtein, D. Ab Initio Molecular Dynamics and

Lattice Dynamics-Based Force Field for Modeling Hexagonal Boron Nitride in Mechanical and Interfacial Applications. *J. Phys. Chem. Lett.* **2018**, *9*, 1584–1591.

(52) Zeron, I. M.; Abascal, J. L. F.; Vega, C. A Force Field of Li$^+$, Na$^+$, K$^+$, Mg$^{2+}$, Ca$^{2+}$, Cl$^-$, and SO$_4^{2-}$ in Aqueous Solution Based on the TIP4P/2005 Water Model and Scaled Charges for the Ions. *J. Chem. Phys.* **2019**, *151*, 134504.

(53) Govind Rajan, A.; Strano, M. S.; Blankschtein, D. Liquids with Lower Wettability Can Exhibit Higher Friction on Hexagonal Boron Nitride: The Intriguing Role of Solid–Liquid Electrostatic Interactions. *Nano Lett.* **2019**, *19*, 1539–1551.

(54) Abal, J. P. K.; Bordin, J. R.; Barbosa, M. C. Salt Parameterization Can Drastically Affect the Results from Classical Atomistic Simulations of Water Desalination by MoS$_2$ Nanopores. *Phys. Chem. Chem. Phys.* **2020**, *22*, 11053–11061.

(55) Kühne, T. D.; Iannuzzi, M.; Del Ben, M.; Rybkin, V. V.; Seewald, P.; Stein, F.; Laino, T.; Khaliullin, R. Z.; Schütt, O.; Schiffmann, F.; Golze, D.; Wilhelm, J.; Chulkov, S.; Bani-Hashemian, M. H.; Weber, V.; Borštnik, U.; Taillefumier, M.; Jakobovits, A. S.; Lazzaro, A.; Pabst, H.; Müller, T.; Schade, R.; Guidon, M.; Andermatt, S.; Holmberg, N.; Schenter, G. K.; Hehn, A.; Bussy, A.; Belleflamme, F.; Tabacchi, G.; Glöß, A.; Lass, M.; Bethune, I.; Mundy, C. J.; Plessl, C.; Watkins, M.; VandeVondele, J.; Krack, M.; Hutter, J. CP2K: An Electronic Structure and Molecular Dynamics Software Package - Quickstep: Efficient and Accurate Electronic Structure Calculations. *J. Chem. Phys.* **2020**, *152*, 194103.

(56) Perdew, J. P.; Burke, K.; Ernzerhof, M. Generalized Gradient Approximation Made Simple. *Phys. Rev. Lett.* **1996**, *77*, 3865–3868.

(57) VandeVondele, J.; Hutter, J. Gaussian Basis Sets for Accurate Calculations on Molecular Systems in Gas and Condensed Phases. *J. Chem. Phys.* **2007**, *127*, 114105.

(58) Goedecker, S.; Teter, M.; Hutter, J. Separable Dual-Space Gaussian Pseudopotentials. *Phys. Rev. B* **1996**, *54*, 1703–1710.

(59) Blöchl, P. E. Electrostatic Decoupling of Periodic Images of Plane-Wave-Expanded Densities and Derived Atomic Point Charges. *J. Chem. Phys.* **1995**, *103*, 7422–7428.

(60) Xu, N.; Guo, J. G.; Cui, Z. The Influence of Tilt Grain Boundaries on the Mechanical Properties of Bicrystalline Graphene Nanoribbons. *Phys. E Low-Dimensional Syst. Nanostructures* **2016**, *84*, 168–174.

(61) Yazyev, O. V.; Louie, S. G. Electronic Transport in Polycrystalline Graphene. *Nat. Mater.* **2010**, *9*, 806–809.

(62) Gibb, A. L.; Alem, N.; Chen, J.-H.; Erickson, K. J.; Ciston, J.; Gautam, A.; Linck, M.; Zettl, A. Atomic Resolution Imaging of Grain Boundary Defects in Monolayer Chemical Vapor Deposition-Grown Hexagonal Boron Nitride. *J. Am. Chem. Soc.* **2013**, *135*, 6758–6761.

(63) Ren, X.; Dong, J.; Yang, P.; Li, J.; Lu, G.; Wu, T.; Wang, H.; Guo, W.; Zhang, Z.; Ding, F.; Jin, C. Grain Boundaries in Chemical-Vapor-Deposited Atomically Thin Hexagonal Boron Nitride. *Phys. Rev. Mater.* **2019**, *3*, 1–11.

(64) Ren, X.; Jin, C. Grain Boundary Motion in Two-Dimensional Hexagonal Boron Nitride. *ACS Nano* **2020**, *14*, 13512–13523.

(65) Li, Q.; Zou, X.; Liu, M.; Sun, J.; Gao, Y.; Qi, Y.; Zhou, X.; Yakobson, B. I.; Zhang, Y.; Liu, Z. Grain Boundary Structures and Electronic Properties of Hexagonal Boron Nitride on Cu(111). *Nano Lett.* **2015**, *15*, 5804–5810.

(66) Zhang, J.; Zhao, J. Structures and Electronic Properties of Symmetric and Nonsymmetric Graphene Grain Boundaries. *Carbon N. Y.* **2013**, *55*, 151–159.

(67) Sharma, B. B.; Parashar, A. Atomistic Simulations to Study the Effect of Grain Boundaries and Hydrogen Functionalization on the Fracture Toughness of Bi-Crystalline h-BN Nanosheets. *Phys. Chem. Chem. Phys.* **2019**, *21*, 13116–13125.

(68) Wei, Y.; Wu, J.; Yin, H.; Shi, X.; Yang, R.; Dresselhaus, M. The Nature of Strength Enhancement and Weakening by Pentagon – Heptagon Defects in Graphene. *Nat. Mater.* **2012**, *11*, 759–763.

(69) Liu, Y.; Zou, X.; Yakobson, B. I. Dislocations and Grain Boundaries in Two-Dimensional Boron Nitride. *ACS Nano* **2012**, *6*, 7053–7058.

(70) Kotakoski, J.; Jin, C. H.; Lehtinen, O.; Suenaga, K.; Krasheninnikov, A. V. Electron Knock-on Damage in Hexagonal Boron Nitride Monolayers. *Phys. Rev. B* **2010**, *82*, 113404.

(71) Ryu, G. H.; Park, H. J.; Ryou, J.; Park, J.; Lee, J.; Kim, G.; Shin, H. S.; Bielawski, C. W.; Ruoff, R. S.; Hong, S.; Lee, Z. Atomic-Scale Dynamics of Triangular Hole Growth in Monolayer Hexagonal Boron Nitride under Electron Irradiation. *Nanoscale* **2015**, *7*, 10600–10605.

(72) Gilbert, S. M.; Dunn, G.; Azizi, A.; Pham, T.; Shevitski, B.; Dimitrov, E.; Liu, S.; Aloni, S.; Zettl, A. Fabrication of Subnanometer-Precision Nanopores in Hexagonal Boron Nitride. *Sci. Rep.* **2017**, *7*, 15096.

(73) Govind Rajan, A.; Silmore, K. S.; Swett, J.; Robertson, A. W.; Warner, J. H.; Blankschtein, D.; Strano, M. S. Addressing the Isomer Cataloguing Problem for Nanopores in Two-Dimensional Materials. *Nat. Mater.* **2019**, *18*, 129–135.

(74) Kozawa, D.; Govind Rajan, A.; Li, S. X.; Ichihara, T.; Koman, V. B.; Zeng, Y.; Kuehne, M.; Iyemperumal, S. K.; Silmore, K. S.; Parviz, D.; Liu, P.; Liu, A. T.; Faucher, S.; Yuan, Z.; Xu, W.; Warner, J. H.; Blankschtein, D.; Strano, M. S. Observation and Spectral Assignment of a Family of Hexagonal Boron Nitride Lattice Defects. **2019**.

(75) Jin, C.; Lin, F.; Suenaga, K.; Iijima, S. Fabrication of a Freestanding Boron Nitride Single Layer and Its Defect Assignments. *Phys. Rev. Lett.* **2009**, *102*, 3–6.

(76) Meyer, J. C.; Chuvilin, A.; Algara-Siller, G.; Biskupek, J.; Kaiser, U. Selective Sputtering and Atomic Resolution Imaging of Atomically Thin Boron Nitride Membranes. *Nano Lett.* **2009**, *9*, 2683–2689.

(77) Yuan, Z.; Govind Rajan, A.; Misra, R. P.; Drahushuk, L. W.; Agrawal, K. V.; Strano, M. S.; Blankschtein, D. Mechanism and Prediction of Gas Permeation through Sub-Nanometer Graphene Pores: Comparison of Theory and Simulation. *ACS Nano* **2017**, *11*, 7974–7987.

(78) Davenport, D. M.; Deshmukh, A.; Werber, J. R.; Elimelech, M. High-Pressure Reverse Osmosis for Energy-Efficient Hypersaline Brine Desalination: Current Status, Design Considerations, and Research Needs. *Environ. Sci. Technol. Lett.* **2018**, *5*, 467–475.

(79) Corry, B. Designing Carbon Nanotube Membranes for Efficient Water Desalination. *J. Phys. Chem. B* **2008**, *112*, 1427–1434.

(80) Shannon, R. D. Revised Effective Ionic Radii and Systematic Studies of Interatomic Distances in Halides and Chalcogenides. *Acta Crystallogr. Sect. A* **1976**, *32*, 751–767.

(81) Werber, J. R.; Deshmukh, A.; Elimelech, M. The Critical Need for Increased Selectivity, Not Increased Water Permeability, for Desalination Membranes. *Environ. Sci. Technol. Lett.* **2016**, *3*, 112–120.

(82) Mulliken, R. S. Electronic Population Analysis on LCAO–MO Molecular Wave Functions. I. *J. Chem. Phys.* **1955**, *23*, 1833–1840.

(83) Hirshfeld, F. L. Bonded-Atom Fragments for Describing Molecular Charge Densities. *Theor. Chim. Acta* **1977**, *44*, 129–138.

(84) Manz, T. A.; Sholl, D. S. Chemically Meaningful Atomic Charges That Reproduce the Electrostatic Potential in Periodic and Nonperiodic Materials. *J. Chem. Theory Comput.* **2010**, *6*, 2455–2468.

(85) Seal, A.; Govind Rajan, A. Modulating Water Slip Using Atomic-Scale Defects: Friction on Realistic Hexagonal Boron Nitride Surfaces. *Nano Lett.* **2021**, *21*, 8008–8016.

**TOC Graphic**

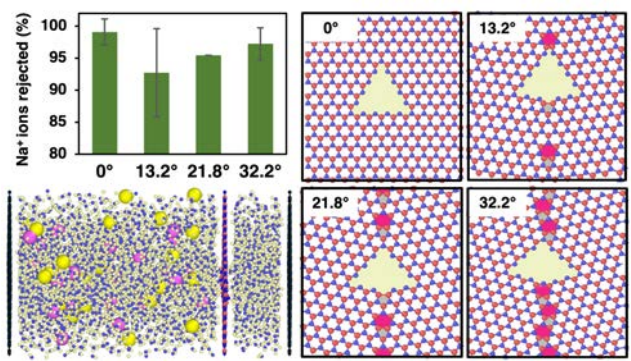

**Supporting Information for:**

*How Grain Boundaries and Interfacial Electrostatic Interactions Affect Water Desalination Via Nanoporous Hexagonal Boron Nitride*

*Bharat Bhushan Sharma and Ananth Govind Rajan*[*]

Department of Chemical Engineering, Indian Institute of Science, Bengaluru, Karnataka 560012, India

**\*Corresponding Author:** Ananth Govind Rajan (Email: ananthgr@iisc.ac.in)

**Table of Contents**

## S1. Water flow rate for nanoporous monocrystalline hBN membrane at 50 MPa feed pressure when the permeate side has 500 and 2000 water molecules

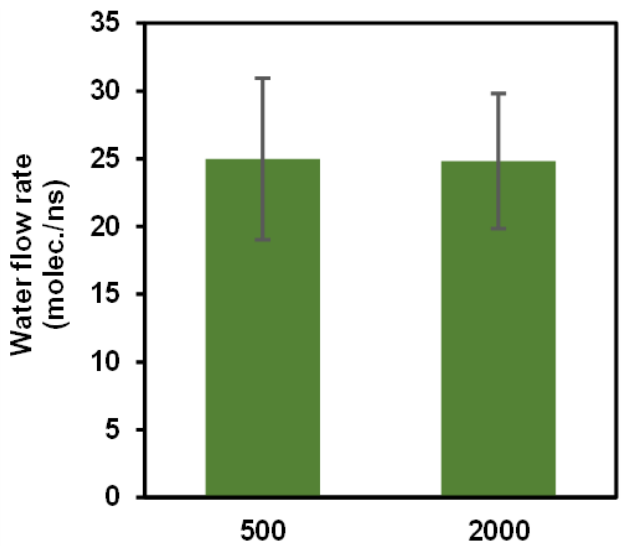

**Figure S1.** Water flow rate for nanoporous monocrystalline hBN membrane at 50 MPa feed pressure by considering 500 and 2000 water molecules on the permeate side.

## S2. Convergence of energy and temperature during equilibration

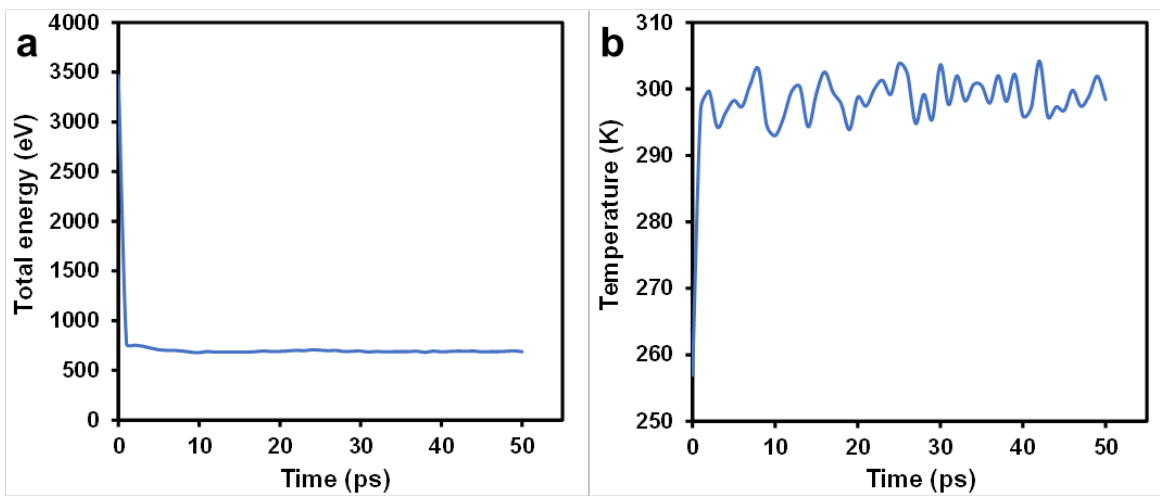

**Figure S2.** Convergence plots during equilibration for the (a) total energy and (b) system temperature.

## S3. Parameters for the LJ and Coulombic interatomic potentials

The parameters for LJ and Coulombic interatomic potentials were adapted from various studies in the literature and are mentioned in Table S1.

**Table S1.** LJ parameters and partial charges on various atoms. Partial charges on B and N atoms are not mentioned here as they were determined separately using density derived atomic point (DDAP) charges based DFT calculations.

| Atom | $\sigma$ (Å) | $\varepsilon$ (kcal/mol) | $q$ (e) | Reference |
|---|---|---|---|---|
| Na$^+$ (Salt) | 2.21737 | 0.35190153 | 0.85 | Zeron et al.[1] |
| Cl$^-$ (Salt) | 4.69906 | 0.01838504 | -0.85 | Zeron et al.[1] |
| C (Piston) | 3.4 | 0.05568834 | 0 | Konatham et al.[2] |
| B (Membrane) | 3.3087 | 0.06924 | - | Govind Rajan et al.[3] |
| N (Membrane) | 3.2174 | 0.047299 | - | Govind Rajan et al.[3] |
| H (Water) | 0 | 0 | 0.5564 | Abascal et al.[4] |
| O (Water) | 0.1852 | 3.1589 | -1.1128 | Abascal et al.[4] |

**S4. Choice of the force fields for hBN, water molecules, salt ions, and interfacial interactions between hBN and water**

Any molecular dynamics (MD) simulation's accuracy is critically dependent upon the chosen force field (FF). For hBN, the classical FF model developed by Govind Rajan et al.,[3] Lennard-Jones (LJ) potential[5] with parameters taken from the DREIDING FF[6] or CHARMM FF,[7] and the Tersoff FF model[8] with parameters suggested by Sevik et al.[9] have been used to simulate the interaction between boron and nitrogen atoms.[10–16] Out of these force fields, the Tersoff type potential is the most suitable force field to simulate grain boundaries in the hBN membrane, due to the presence of B-B/N-N bonds, and has been used by several researchers.[17–19] In this work, the Tersoff parameters suggested by Albe et al.[20] were used. In addition to the hBN membrane, water molecules need to be modelled accurately. For this purpose, several water models, e.g., SPC,[21] SPC/E,[22] SPC/Fw,[23] TIP3P,[24] TIP4P,[25] TIP4P/2005,[4] etc. are available to simulate the water molecules. In 2018, Prasad et al.[26] performed MD simulations to investigate the influence of the water model on water desalination using graphene nanopores and predicted that the calculated water flux varied up to 84% between the water models. The authors suggested the use of the TIP4P/2005 water model for computational studies on desalination across nanoporous membranes because the calculated diffusion coefficient and shear viscosity of water using this model are very close to the respective experimental values. Thus, in this work, we have used the TIP4P/2005 water model to simulate water molecules. Apart from this, the LJ parameters and partial charges used for $Na^+$ and $Cl^-$ ions also affect the water permeability, especially for narrow pore.[27] Although, several LJ parameters[1,28,29] have been used to simulate the $Na^+$ and $Cl^-$ ions in water desalination, we used the LJ parameters proposed by Zeron et al.[1] since they very well reproduce the density and visocity of salt-water solutions across a range of concentrations.[1]

**S5. Model used to calculate the interfacial interactions between a nanoporous hBN membrane and a unit test charge/tangential water molecule**

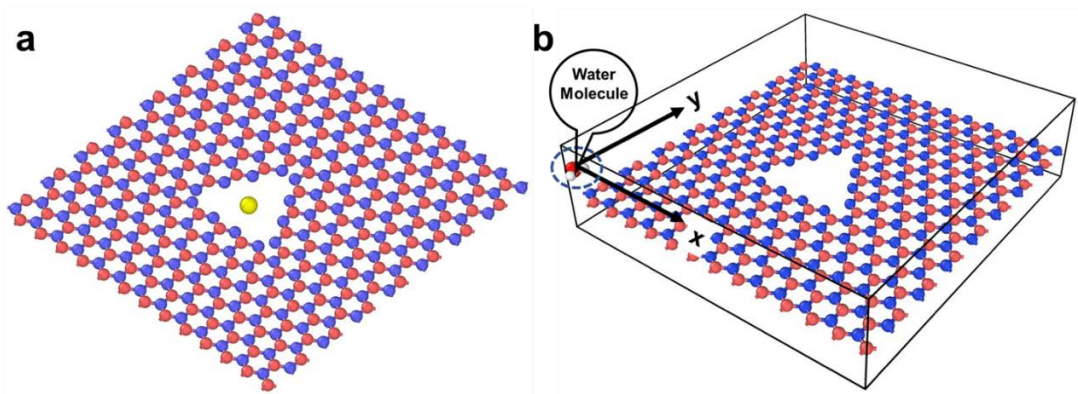

**Figure S3.** Snapshot of a nanoporous hBN membrane (boron in red and nitrogen in blue) depicting (a) a test charge atom (in yellow) at the centre of the nanopore and (b) a tangential water molecule at the corner of the membrane at a distance of 3 Å above the surface of the hBN layer.

## S6. Nomenclature and repeat length of the grain boundaries (GBs)

The nomenclature of a GB is illustrated in Figure S4. A GB can be described by its misorientation angle ($\theta = \theta_L + \theta_R$), where $\theta_L$ and $\theta_R$ represent the rotation angles of the left and right crystals, respectively. Note that, for symmetric tilt GBs, $\theta_L = \theta_R = \frac{\theta}{2}$.[30]

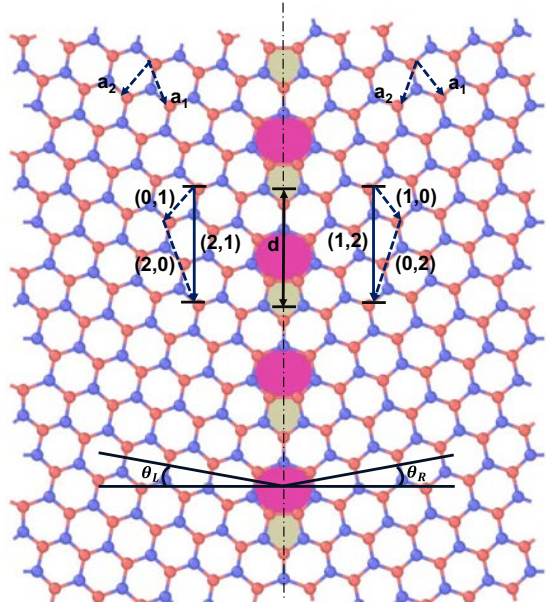

**Figure S4.** The structure of symmetric tilt GB in hBN having misorientation angle of 21.8°. The misorientation angle ($\theta = \theta_L + \theta_R = 10.9° + 10.9° = 21.8°$) described the misalignment between the two crystals. The vectors (2,1) and (1,2) represent the two periodic translation vectors for the left crystal ($n_L$, $m_L$) and for the right crystal ($n_R$, $m_R$) along the GB. The repeat vector or periodic length (d) of the GB described by the periodic translation vectors at the left and right crystals, respectively.

The GBs in hBN have periodically arranged defects (dislocations (5|7 pairs)), as shown in Figure S4. In the case of periodically arranged defects, an appropriate way to describe the GB is using two periodic translation vectors each for the left crystal ($n_L$, $m_L$) and for the right crystal ($n_R$, $m_R$) of the GB along the defect direction, i.e., as ($n_L$, $m_L$)|($n_R$, $m_R$), which is shown in Figure S4.[30,31] The misorientation angle and periodic translation vectors are related by the following formula: $\theta = \tan^{-1}[\sqrt{3}\, m_L/(m_L + 2n_L)] + \tan^{-1}[\sqrt{3}\, m_R/(m_R + 2n_R)]$.[32] The periodic (i.e., repeat) length for the symmetric tilt GBs, $d$, is calculated using $d = a_0\sqrt{n_L^2 + n_L m_L + m_L^2} = a_0\sqrt{n_R^2 + n_R m_R + m_R^2}$,

where $a_0 = a_1 = a_2 = 2.5115$ Å is the unit vector length of the hBN lattice.[31] The periodic translation vectors and the repeat length corresponding to the three considered misorientation angles, i.e., 13.2°, 21.8°, and 31.8° are (3,2)|(2,3), (2,1)|(1,2), and (3,1)|(1,3) and 10.9474 Å, 6.6448 Å, and 9.0553 Å, respectively.

**S7. Calculation of the areas, equivalent circular diameters, and aspect ratios of the various nanopores considered**

In each case, the nanopore area was estimated by considering the B and N atoms as discs of radius equal to the van der Waals radius of the respective atom[33] (1.92 Å for B and 1.55 Å for N) in the plane of the membrane (Figure S5). To this end, a code written in MATLAB R2021a was used, whereby the number of pixels occupied by the nanopore, after excluding the B and N discs, were used to determine the area, $A$, of the nanopore using the MATLAB functions bwconncomp and regionprops. A similar approach has been used in other studies to determine the nanopore area.[34,35] Once the area of the nanopore was determined, the equivalent circular diameter, $d$, of the nanopore, was calculated as $d = \sqrt{\frac{4A}{\pi}}$. The maximum ($d_{F,max}$) and minimum ($d_{F,min}$) Feret diameters of each nanopore were also determined using the regionprops function in MATLAB (Figure S5). Subsequently, the aspect ratio, $S$, was calculated as $S = \frac{d_{F,min}}{d_{F,max}}$.

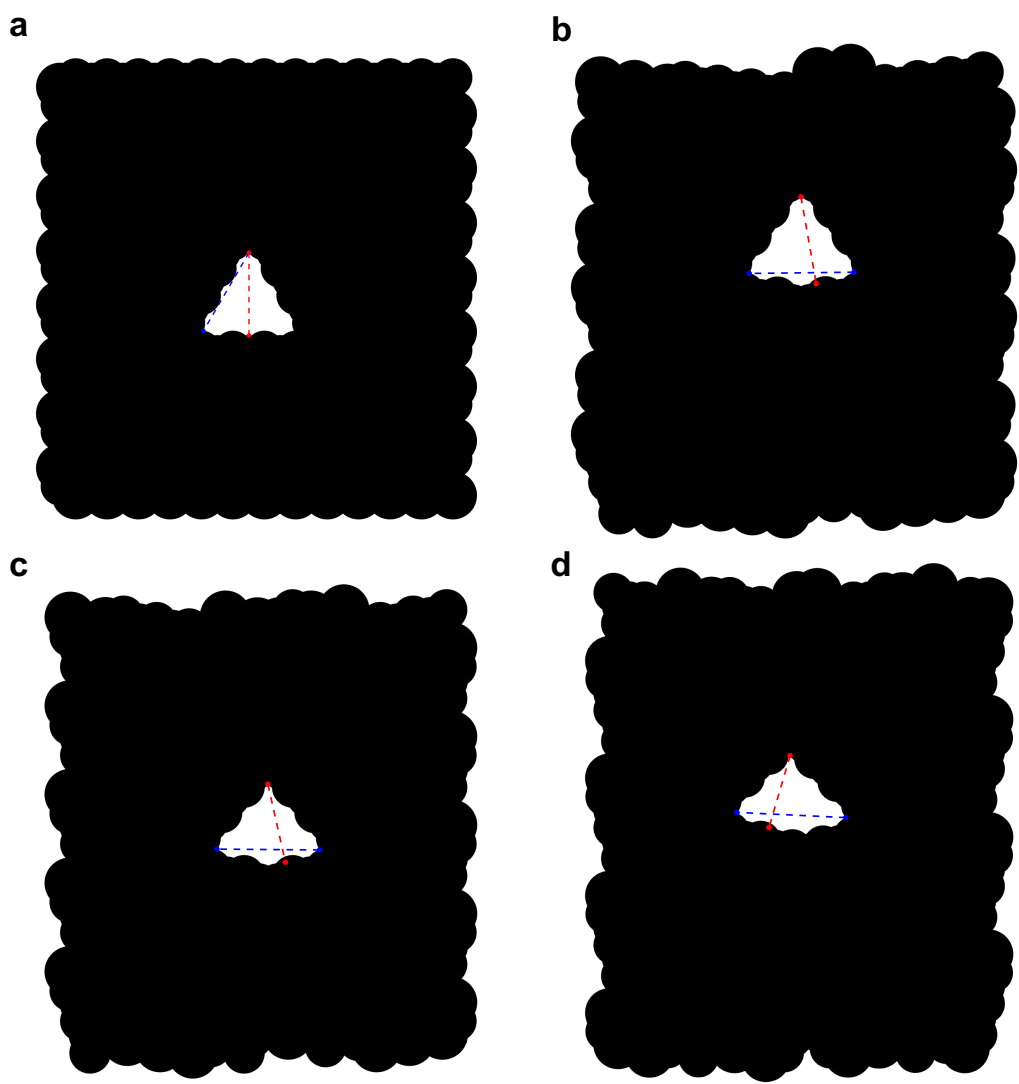

**Figure S5.** Calculation of the nanopore area for (a) monocrystalline hBN, (b) 13.2°-misoriented bicrystalline hBN, (c) 21.8°-misoriented bicrystalline hBN, and (d) 32.2°-misoriented bicrystalline hBN. The dashed red line indicates the minimum Feret diameter and the dashed blue line indicates the maximum Feret diameter.

## S8. Tuning of partial charges and validation of the resultant electrostatic potential for 13.2°-misoriented bicrystalline nanoporous hBN

To eliminate the differences between the electrostatic potentials obtained in molecular dynamics (MD) simulations and density functional theory (DFT) calculations, we tuned the partial charges by altering the parameters for the density derived atomic point (DDAP) charges method.[36] The nine sets of parameters that were investigated are tabulated in Table S2. Note that the parameters denoted as "case 1" are the default parameters for the DDAP method in the cp2k package. For each case, the partial charges on B and N atoms were evaluated, and the corresponding electrostatic potentials were determined using single-point calculations in LAMMPS and plotted in Figure S6. Apart from visually comparing the partial-charge-based electrostatic potential with the DFT-based potential in Figure S6, the root-mean square deviation (RMSD) between the two cases was also calculated for each set of parameters and tabulated in Table S2. It can be inferred from the RMSD values that the set of parameters described in case 4 ("best" parameters) shows the minimum deviation, and the calculated partial charges corresponding to these parameters were used for investigating the desalination performance of 13.2°-misoriented bicrystalline nanoporous hBN. The same set of customized parameters, as described in case 4, was also used for obtaining the partial charges for B and N atoms in monocrystalline and bicrystalline (21.8°- and 32.2°-misoriented) nanoporous hBN membranes.

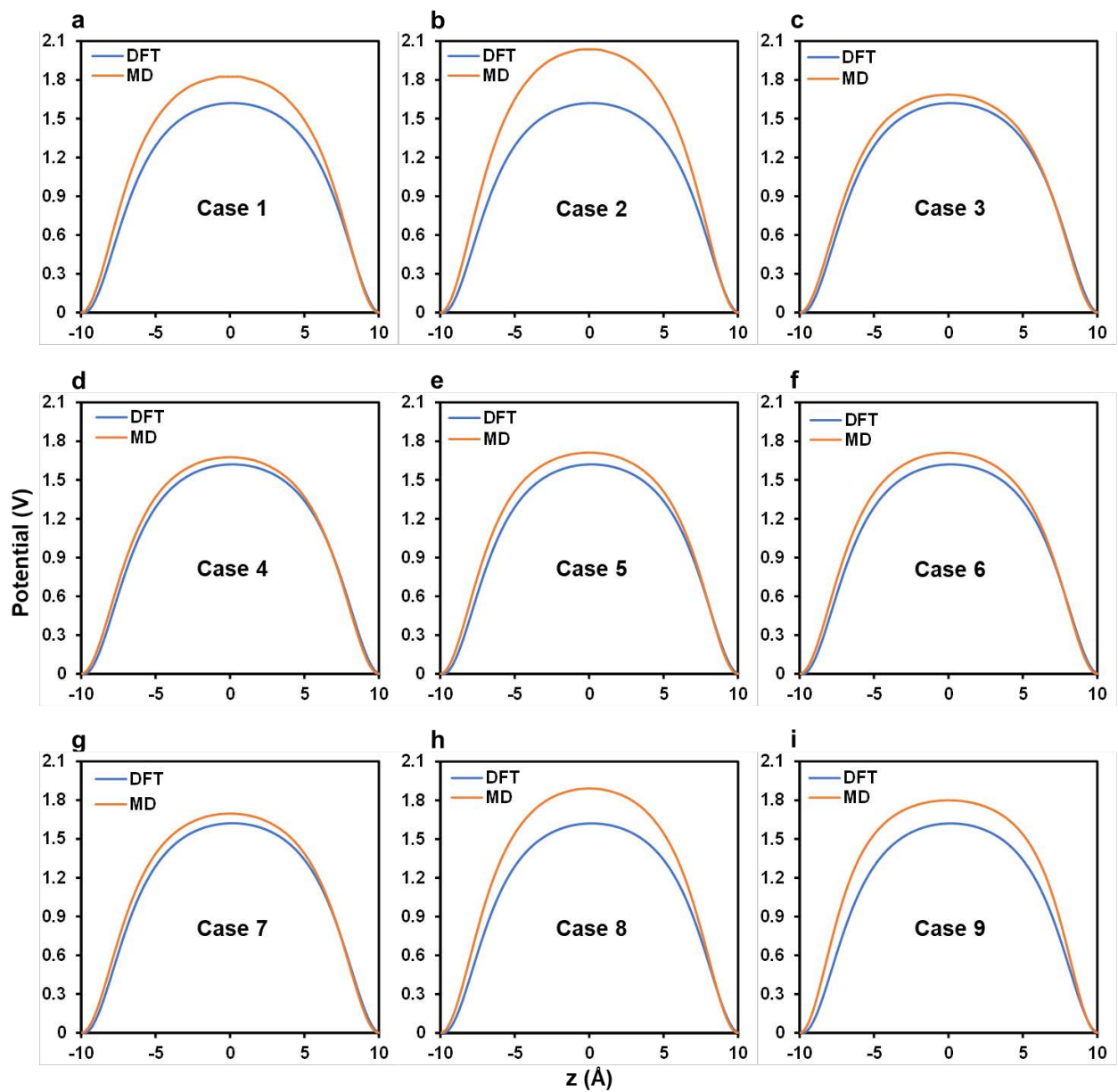

**Figure S6.** Electrostatic potential (in orange color) as a function of the out-of-plane distance from the centre of a 13.2°-misoriented bicrystalline hBN membrane for different sets of parameters using density derived atomic point charges: (a-i) cases 1 to 9 for different set of parameters, respectively. The DFT potential is shown in purple color in each case.

**Table S2.** Different sets of parameters investigated for obtaining the DDAP charges and their resultant RMSD between the DFT-predicted and MD-predicted electric potentials.

| Case | GCUT | NUM_GAUSS | PFACTOR | MIN_RADIUS | RMSD |
|---|---|---|---|---|---|
| 1 | 2.449 | 3 | 1.5 | 0.264589 | 0.16284174 |
| 2 | 5 | 3 | 1.5 | 0.264589 | 0.31203197 |
| 3 | 2.449 | 4 | 1.31 | 0.264589 | 0.06291837 |
| 4 | 2.449 | 4 | 1.31 | 0.3625 | 0.05630271 |
| 5 | 2.449 | 4 | 1.31 | 0.529177 | 0.08840776 |
| 6 | 2.449 | 4 | 1.5 | 0.3625 | 0.08096275 |
| 7 | 2.449 | 4 | 1.7 | 0.264589 | 0.07029125 |
| 8 | 5 | 4 | 1.5 | 0.264589 | 0.21126815 |
| 9 | 5 | 4 | 1.5 | 0.3625 | 0.18728606 |

**S9. Comparing two different methods for water permeability calculation**

In this work, the water permeability was calculated using two methods: (i) the total number of water molecules that permeated through the membrane divided by the time taken, (ii) by obtaining the slope of a straight line fitted to the $N_w$ versus time plot. The water permeabilities of nanoporous monocrystalline hBN membrane determined with the help of above-mentioned methods were plotted against the effective pressure in Figure S7a. In addition, water permeabilities of nanoporous monocrystalline hBN membrane were determined when the hBN membrane is simulated using the interatomic potential developed by Govind Rajan et al.[3] and plotted against the effective pressure in Figure S7b. It can be inferred from Figure S7 that the water permeabilities calculated by the two methods closely agree with each other at every value of the effective pressure, while using any of the two intra-hBN potentials. In other words, straight-line fitting gives similar results as monitoring the total number of permeated molecules in a long-time simulation.

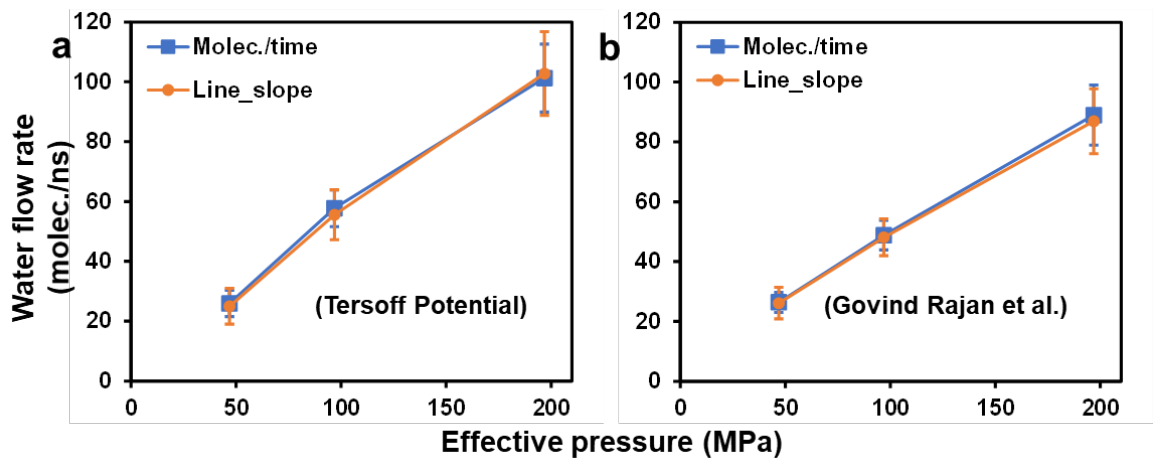

**Figure S7.** Water permeability versus effective pressure plots, calculated using two different methods, for a nanoporous monocrystalline hBN membrane when the membrane was simulated using: (a) the Tersoff force field[20] and (b) the force field developed by Govind Rajan et al.[3]

**S10. Comparison of the classical electrostatic potentials calculated using DDAP, Hirshfeld, Mulliken, and density derived electrostatic and chemical (DDEC) partial charges with the DFT-derived potential**

To determine the algorithm for calculating partial charges that is best-suited for predicting the electrostatic potential above a nanopore in hBN, we compared the classical electrostatic potential calculated using DDAP, Hirshfeld, Mulliken, and DDEC partial charges with the DFT-derived potential in Figure S8. It can be inferred from Figure S8 that the electrostatic potential calculated using DDAP partial charges with customized parameters (case 4) shows the best agreement with the DFT-derived potential, whereas the electrostatic potentials calculated using the default DDAP, Hirshfeld, Mulliken, and DDEC partial charges show notable deviations. Further, the deviation in the case of Mulliken charges is more significant than in the case of other charges.

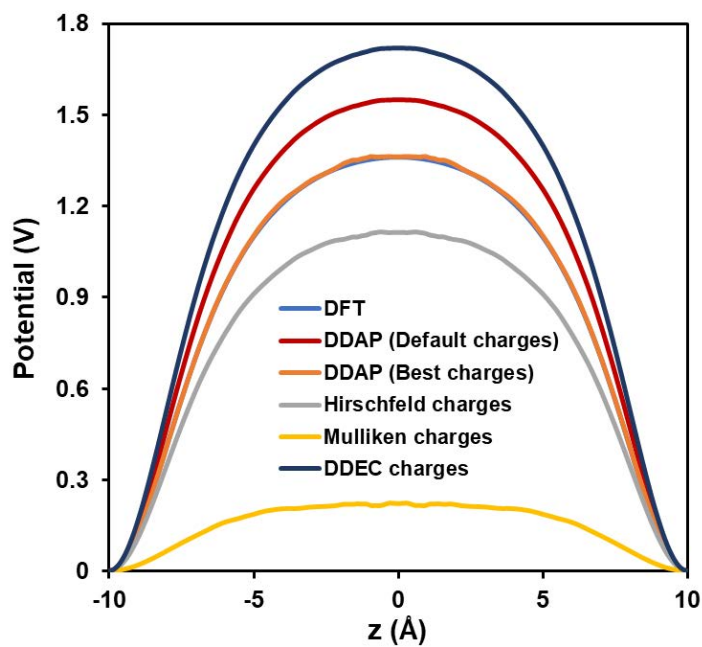

**Figure S8.** Electrostatic potential calculated using DDAP partial charges with default and customized parameters, Hirshfeld partial charges, Mulliken partial charges, DDEC partial charges, and direct DFT calculation for a monocrystalline nanoporous hBN membrane.

## S11. References

(1) Zeron, I. M.; Abascal, J. L. F.; Vega, C. A Force Field of Li$^+$ , Na$^+$ , K$^+$ , Mg$^{2+}$ , Ca$^{2+}$ , Cl$^-$, and SO$_4^{2-}$ in Aqueous Solution Based on the TIP4P/2005 Water Model and Scaled Charges for the Ions. *J. Chem. Phys.* **2019**, *151* (13), 134504. https://doi.org/10.1063/1.5121392.

(2) Konatham, D.; Yu, J.; Ho, T. A.; Striolo, A. Simulation Insights for Graphene-Based Water Desalination Membranes. *Langmuir* **2013**, *29* (38), 11884–11897. https://doi.org/10.1021/la4018695.

(3) Govind Rajan, A.; Strano, M. S.; Blankschtein, D. Ab Initio Molecular Dynamics and Lattice Dynamics-Based Force Field for Modeling Hexagonal Boron Nitride in Mechanical and Interfacial Applications. *J. Phys. Chem. Lett.* **2018**, *9* (7), 1584–1591. https://doi.org/10.1021/acs.jpclett.7b03443.

(4) Abascal, J. L. F.; Vega, C. A General Purpose Model for the Condensed Phases of Water: TIP4P/2005. *J. Chem. Phys.* **2005**, *123* (23), 234505. https://doi.org/10.1063/1.2121687.

(5) Jones, J. E. On the Determination of Molecular Fields. —II. From the Equation of State of a Gas. *Proc. R. Soc. London. Ser. A, Contain. Pap. a Math. Phys. Character* **1924**, *106* (738), 463–477. https://doi.org/10.1098/rspa.1924.0082.

(6) Mayo, S. L.; Olafson, B. D.; Goddard, W. A. DREIDING: A Generic Force Field for

Molecular Simulations. *J. Phys. Chem.* **1990**, *94* (26), 8897–8909. https://doi.org/10.1021/j100389a010.

(7) Brooks, B. R.; Bruccoleri, R. E.; Olafson, B. D.; States, D. J.; Swaminathan, S.; Karplus, M. CHARMM: A Program for Macromolecular Energy, Minimization, and Dynamics Calculations. *J. Comput. Chem.* **1983**, *4* (2), 187–217. https://doi.org/10.1002/jcc.540040211.

(8) Tersoff, J. New Empirical Approach for the Structure and Energy of Covalent Systems. *Phys. Rev. B* **1988**, *37* (12), 6991–7000. https://doi.org/10.1103/PhysRevB.37.6991.

(9) Kınacı, A.; Haskins, J. B.; Sevik, C.; Çağın, T. Thermal Conductivity of BN-C Nanostructures. *Phys. Rev. B* **2012**, *86* (11), 115410. https://doi.org/10.1103/PhysRevB.86.115410.

(10) Tsukanov, A. A.; Shilko, E. V. Computer-Aided Design of Boron Nitride-Based Membranes with Armchair and Zigzag Nanopores for Efficient Water Desalination. *Materials (Basel).* **2020**, *13* (22), 1–12. https://doi.org/10.3390/ma13225256.

(11) Liu, L.; Liu, Y.; Qi, Y.; Song, M.; Jiang, L.; Fu, G.; Li, J. Hexagonal Boron Nitride with Nanoslits as a Membrane for Water Desalination: A Molecular Dynamics Investigation. *Sep. Purif. Technol.* **2020**, *251*, 117409. https://doi.org/10.1016/j.seppur.2020.117409.

(12) Srivastava, R.; Kommu, A.; Sinha, N.; Singh, J. K. Removal of Arsenic Ions Using Hexagonal Boron Nitride and Graphene Nanosheets: A Molecular Dynamics Study. *Mol. Simul.* **2017**, *43* (13–16), 985–996. https://doi.org/10.1080/08927022.2017.1321754.

(13) Gao, H.; Shi, Q.; Rao, D.; Zhang, Y.; Su, J.; Liu, Y.; Wang, Y.; Deng, K.; Lu, R. Rational Design and Strain Engineering of Nanoporous Boron Nitride Nanosheet Membranes for Water Desalination. *J. Phys. Chem. C* **2017**, *121* (40), 22105–22113. https://doi.org/10.1021/acs.jpcc.7b06480.

(14) Garnier, L.; Szymczyk, A.; Malfreyt, P.; Ghoufi, A. Physics behind Water Transport through Nanoporous Boron Nitride and Graphene. *J. Phys. Chem. Lett.* **2016**, *7* (17), 3371–3376. https://doi.org/10.1021/acs.jpclett.6b01365.

(15) Nasser Saadat Tabrizi ; Behrouz Vahid ;Jafar Azamat. Functionalized Single-Atom Thickness Boron Nitride Membrane for Separation of Arsenite Ion from Water: A Molecular Dynamics Simulation Study. **2020**, *8* (3), 843–856. https://doi.org/10.22036/PCR.2020.222756.1742.

(16) Loh, G. C. Fast Water Desalination by Carbon-Doped Boron Nitride Monolayer: Transport Assisted by Water Clustering at Pores. *Nanotechnology* **2019**, *30* (5), 055401. https://doi.org/10.1088/1361-6528/aaf063.

(17) Li, Y.; Wei, A.; Ye, H.; Yao, H. Mechanical and Thermal Properties of Grain Boundary in a Planar Heterostructure of Graphene and Hexagonal Boron Nitride. *Nanoscale* **2018**, *10* (7), 3497–3508. https://doi.org/10.1039/C7NR07306B.

(18) Sharma, B. B.; Parashar, A. Inter-Granular Fracture Behaviour in Bicrystalline Boron Nitride Nanosheets Using Atomistic and Continuum Mechanics-Based Approaches. *J. Mater. Sci.* **2021**, *56* (10), 6235–6250. https://doi.org/10.1007/s10853-020-05697-x.

(19) Ding, Q.; Ding, N.; Liu, L.; Li, N.; Wu, C.-M. L. Investigation on Mechanical Performances of Grain Boundaries in Hexagonal Boron Nitride Sheets. *Int. J. Mech. Sci.* **2018**, *149* (October), 262–272. https://doi.org/10.1016/j.ijmecsci.2018.10.003.

(20) Albe, K.; Möller, W.; Heinig, K. Computer Simulation and Boron Nitride. *Radiat. Eff. Defects Solids* **1997**, *141*, 85–97. https://doi.org/10.1080/10420159708211560.

(21) Berendsen, H. J. C.; Postma, J. P. M.; van Gunsteren, W. F.; Hermans, J. Interaction Models for Water in Relation to Protein Hydration; 1981; pp 331–342. https://doi.org/10.1007/978-94-015-7658-1_21.

(22) Berendsen, H. J. C.; Grigera, J. R.; Straatsma, T. P. The Missing Term in Effective Pair Potentials. *J. Phys. Chem.* **1987**, *91* (24), 6269–6271. https://doi.org/10.1021/j100308a038.

(23) Wu, Y.; Tepper, H. L.; Voth, G. A. Flexible Simple Point-Charge Water Model with Improved Liquid-State Properties. *J. Chem. Phys.* **2006**, *124* (2), 024503. https://doi.org/10.1063/1.2136877.

(24) Price, D. J.; Brooks, C. L. A Modified TIP3P Water Potential for Simulation with Ewald Summation. *J. Chem. Phys.* **2004**, *121* (20), 10096–10103. https://doi.org/10.1063/1.1808117.

(25) William L. Jorgensen, Jayaraman Chandrasekhar, and J. D. M. Comparison of Simple Potential Functions for Simulating Liquid Water. *J. Chem. Phys.* **1983**, *79* (2), 926. https://doi.org/10.1063/1.445869.

(26) Prasad K, V.; Kannam, S. K.; Hartkamp, R.; Sathian, S. P. Water Desalination Using Graphene Nanopores: Influence of the Water Models Used in Simulations. *Phys. Chem. Chem. Phys.* **2018**, *20* (23), 16005–16011. https://doi.org/10.1039/c8cp00919h.

(27) Abal, J. P. K.; Bordin, J. R.; Barbosa, M. C. Salt Parameterization Can Drastically Affect the Results from Classical Atomistic Simulations of Water Desalination by $MoS_2$ Nanopores. *Phys. Chem. Chem. Phys.* **2020**, *22* (19), 11053–11061. https://doi.org/10.1039/D0CP00484G.

(28) Fuentes-Azcatl, R.; Barbosa, M. C. Sodium Chloride, NaCl/ϵ: New Force Field. *J. Phys. Chem. B* **2016**, *120* (9), 2460–2470. https://doi.org/10.1021/acs.jpcb.5b12584.

(29) Joung, I. S.; Cheatham, T. E. Determination of Alkali and Halide Monovalent Ion Parameters for Use in Explicitly Solvated Biomolecular Simulations. *J. Phys. Chem. B* **2008**, *112* (30), 9020–9041. https://doi.org/10.1021/jp8001614.

(30) Zhang, J.; Zhao, J. Structures and Electronic Properties of Symmetric and Nonsymmetric

Graphene Grain Boundaries. *Carbon N. Y.* **2013**, *55*, 151–159. https://doi.org/10.1016/j.carbon.2012.12.021.

(31) Yazyev, O. V.; Louie, S. G. Electronic Transport in Polycrystalline Graphene. *Nat. Mater.* **2010**, *9* (10), 806–809. https://doi.org/10.1038/nmat2830.

(32) Ding, Q.; Ding, N.; Liu, L.; Li, N.; Wu, C.-M. L. Investigation on Mechanical Performances of Grain Boundaries in Hexagonal Boron Nitride Sheets. *Int. J. Mech. Sci.* **2018**, *149*, 262–272. https://doi.org/10.1016/j.ijmecsci.2018.10.003.

(33) Mantina, M.; Chamberlin, A. C.; Valero, R.; Cramer, C. J.; Truhlar, D. G. Consistent van Der Waals Radii for the Whole Main Group. *J. Phys. Chem. A* **2009**, *113* (19), 5806–5812. https://doi.org/10.1021/jp8111556.

(34) Cohen-Tanugi, D.; Grossman, J. C. Water Desalination across Nanoporous Graphene. *Nano Lett.* **2012**, *12* (7), 3602–3608. https://doi.org/10.1021/nl3012853.

(35) Heiranian, M.; Farimani, A. B.; Aluru, N. R. Water Desalination with a Single-Layer $MoS_2$ Nanopore. *Nat. Commun.* **2015**, *6* (1), 8616. https://doi.org/10.1038/ncomms9616.

(36) Blöchl, P. E. Electrostatic Decoupling of Periodic Images of Plane-Wave-Expanded Densities and Derived Atomic Point Charges. *J. Chem. Phys.* **1995**, *103* (17), 7422–7428. https://doi.org/10.1063/1.470314.